# Mechanical Characterization of Brain Tissue: Experimental Techniques, Human Testing Considerations, and Perspectives

Jixin Hou[1], Kun Jiang[1], Arunachalam Ramanathan[1], Abhishek Saji Kumar[2], Wei Zhang[3], Lin Zhao[4], Taotao Wu[5], Ramana Pidaparti[1], Dajiang Zhu[6], Gang Li[7], Kenan Song[1], Tianming Liu[4], Mir Jalil Razavi[8], Ellen Kuhl[9], Xianqiao Wang[1]*

[1]School of Environmental, Civil, Agricultural and Mechanical Engineering, College of Engineering, University of Georgia, Athens, GA 30602, USA

[2]Department of Material Science, School for Engineering of Matter, Transport and Energy, Ira A. Fulton Schools of Engineering, Arizona State University, Tempe, AZ 85287, USA

[3]School of Computer and Cyber Sciences, Augusta University, GA 30602, USA

[4]School of Computing, The University of Georgia, Athens, GA 30602, USA

[5]School of Chemical, Materials, and Biomedical Engineering, College of Engineering, University of Georgia, Athens, GA 30602, USA

[6]Department of Computer Science and Engineering, The University of Texas at Arlington, Arlington, TX 76019, USA

[7]Department of Radiology and Biomedical Research Imaging Center, The University of North Carolina at Chapel Hill, NC 27599, USA

[8]Department of Mechanical Engineering, SUNY Binghamton University, Binghamton, NY 13902, USA

[9]Department of Mechanical Engineering, Stanford University, Stanford, CA 94305, USA

*Corresponding Author: xqwang@uga.edu

# Table of Contents

**Abstract**

Understanding the mechanical behavior of brain tissue is crucial for advancing both fundamental neuroscience and clinical applications. Yet, accurately measuring these properties remains challenging due to the brain's unique mechanical attributes and complex anatomical structures. This review provides a comprehensive overview of commonly used techniques for characterizing brain tissue mechanical properties, covering both invasive methods—such as atomic force microscopy, indentation, axial mechanical testing, and oscillatory shear testing—and noninvasive approaches like magnetic resonance elastography and ultrasound elastography. Each technique is evaluated in terms of working principles, applicability, representative studies, and experimental limitations. We further summarize existing publications that have used these techniques to measure human brain tissue mechanical properties. With a primary focus on invasive studies, we systematically compare their sample preparation, testing conditions, reported mechanical parameters, and modeling strategies. Key sensitivity factors influencing testing outcomes (e.g., sample size, anatomical location, strain rate, temperature, conditioning, and post-mortem interval) are also discussed. Additionally, selected noninvasive studies are reviewed to assess their potential for *in vivo* characterization. A comparative discussion between invasive and noninvasive methods, as well as *in vivo* versus *ex vivo* testing, is included. This review aims to offer practical guidance for researchers and clinicians in selecting appropriate mechanical testing approaches and contributes a curated dataset to support constitutive modeling of human brain tissue.

## 1. Introduction

As the central regulator of the human body, the brain orchestrates a wide range of vital physiological and cognitive functions. Accordingly, brain research spans multiple disciplines, including molecular biology, cellular neuroscience, bioelectrical signaling, and functional imaging. Among these, biomechanics plays a critical yet often underappreciated role. Understanding the brain's mechanical behavior is essential for uncovering fundamental physiological and pathological processes, such as cortical folding during brain development [1-6], traumatic brain injury (TBI) [7, 8], and neurological disease progression [9, 10]. For example, studies have shown that cortical folding arises from mechanical buckling, driven by compressive forces generated through differential growth between gray and white matter [11-13]. In the case of TBI, external impacts induce rapid and excessive shear deformation, leading to immediate tissue damage and long-term degeneration [14-16]. Similarly, neurodegenerative diseases such as Alzheimer's disease (AD) involve progressive tissue degradation, often initiated by aging-related mechanical changes or the spread of toxic proteins [10, 17]. Beyond its role in mechanical understanding, biomechanics also holds increasing promise in brain disorders diagnosis. Variations in tissue stiffness have been correlated with pathological conditions such as brain tumors [10, 17], epilepsy [18, 19], and dementia [20], which offers opportunities for noninvasive disease detection and monitoring. Accurate characterization of brain mechanical properties is therefore indispensable for effectively analyzing the underlying mechanics of these complex phenomena and supporting clinical applications.

Mechanical testing of brain tissue, however, presents significant challenges due to the tissue's complex mechanical characteristics. Brain tissue is ultrasoft, fragile, biphasic, and exhibits pronounced anatomical and microstructural heterogeneity [21]. These attributes complicate both sample preparation and experimental execution. For instance, its fragility constrains the range of applicable deformation to preserve tissue integrity during tests [22, 23]. Anatomical variability restricts consistent sampling, while the ultrasoft nature and potential dehydration of fluidic components can cause dimensional change under the tissue's weight [24-26]. Over the past decades, a variety of testing techniques have been developed to assess brain mechanics at different spatial and temporal scales. These techniques ensure diverse characterizations in brain tissue tailored to specific research objectives. Atomic force microscopy (AFM), for example, enables the measurement of cellular and subcellular mechanical properties, thereby facilitating the investigation of the microstructural relevance to macroscale brain properties [27]. Indentation (IND) offers a versatile platform for probing brain mechanical properties, enabling the assessment of spatially resolved modulus and time-dependent viscoelastic behaviors [28]. Oscillatory shear testing (OST) allows for the evaluation of frequency-dependent viscoelastic properties, aiding the study of the underlying biomechanism in TBI [29]. Meanwhile, continuous stress-strain data collected through axial mechanical testing (AMT) support the development of hyperelastic constitutive models [30], which are essential for simulating convoluted physiological phenomena such as cortical folding during brain development [12, 31, 32]. Despite these achievements, reported mechanical parameters vary widely across studies—often differing by several orders of magnitude—posing significant barriers to both inter- and intra-study comparisons of brain tissue mechanics.

Due to ethical limitations and logistical constraints on human brain experimentation, animal models have been extensively employed to study brain mechanics [33]. Brains from species such as rodents [34, 35], pigs [36, 37], and bovines [38, 39] are often employed as surrogates for the human brain. However, growing evidence indicates notable interspecies differences, not only in anatomical structure but also in mechanical behavior [40]. Variations in the mechanical properties of gray and white matter, strain-rate sensitivity, and regional stiffness patterns can differ remarkably across species [41, 42]. These unignorable discrepancies raise important concerns about the validity of directly translating findings from animal models to humans. The emergence of noninvasive techniques such as magnetic resonance elastography (MRE) and ultrasound elastography (USE) has enabled direct measurement of human brain mechanical properties *in vivo* [43]. These approaches support population-level studies and facilitate statistically robust investigations into how mechanical properties vary with age, gender, and disease [43]. Their noninvasive nature also allows for repeated and continuous measurements of the same individuals over time [44]. Despite these advantages, current noninvasive methods are limited to capturing relatively simple mechanical quantities—such as shear stiffness, storage, and loss moduli—within small deformation ranges to ensure participant safety and comfort. In addition, the shear-related properties derived from these techniques often show noticeable discrepancies compared to those obtained through invasive approaches. This inconsistency naturally raises concerns regarding the comparability and reliability of the reported mechanical data. More broadly, it points to a longstanding issue in brain testing: the divergence of testing outcomes obtained under different experimental conditions, including *in vivo*, *ex vivo*, *in vitro*, and *in situ* settings [45].

In this review, we aim to provide a comprehensive summary of the current state of brain tissue mechanical testing. We begin by introducing six widely used experimental techniques, including AFM, IND, AMT, OST, MRE, and USE. Each method is summarized in terms of its working principles, measurable mechanical parameters, advantages, and limitations, as well as representative studies. Next, we collect and analyze existing data on human brain mechanics from peer-reviewed studies, categorizing them based on whether the methods are invasive or noninvasive, and discussing key sensitivity factors that influence the testing outcomes. Finally, we provide a comparative discussion between invasive and noninvasive techniques, as well as *in vivo* versus *ex vivo* testing. The review is structured as follows: Section 2 introduces the experimental techniques; Section 3 summarizes human brain tissue mechanical data from the literature; and Section 4 concludes with key insights and perspectives for future research on brain tissue mechanical testing and characterization.

## 2. Established Techniques for Quantifying Brain Tissue Mechanical Properties

In this section, we introduce various testing techniques commonly used for characterizing the mechanical properties of soft tissue, with a focus on brain tissue. These include invasive methods such as AFM, IND, AMT, OST, and noninvasive approaches like MRE and USE. In addition to these six primary techniques, other methods—such as pipette aspiration [46, 47], needle-induced cavitation [48-50], and optical-based diffusion correlation spectroscopy [51]—have also been employed to characterize the mechanical properties of brain tissue. Although these approaches show promise in capturing various aspects of brain mechanics, they have yet to gain widespread attention in the field. Each of the six techniques considered operates based on distinct principles and is suited for measuring various mechanical properties across different length scales (from the cellular to organ level) and time scales (from quasistatic to high-rate dynamics), as shown in Figure 1. However, their testing outputs may be biased due to varied sensitivity factors, thereby requiring careful consideration during testing. As a result, each technique has specific advantages and limitations in measuring the mechanical properties of brain tissue. To provide context, we briefly review these aspects based on existing literature. Since this review focuses on brain tissue mechanics, readers interested in a broader review of these techniques for other tissues may refer to Bejgam, et al. [52], Song, et al. [53], and Navindaran, et al. [54].

### 2.1. Atomic force microscopy: brain characterization at cellular and subcellular scale

AFM is a powerful technique for measuring the mechanical properties of brain tissue at the micro- and nanoscale. It operates by detecting the contact interaction between an indenter tip and the tissue surface. As illustrated in Figure 2, an AFM system consists of four main components: a cantilever with an integrated indenter tip to establish contact; a laser beam directed onto the cantilever tip; a position-sensitive photodiode that detects the reflected laser beam to measure cantilever deflection; and a piezo scanner that controls sample movement [27]. Through precise feedback control, AFM functions not only as an imaging tool to generate high-resolution topology images [55, 56], but also as a highly sensitive mechanical measurement system, capable of recording force-displacement curves with piconewton-scale sensitivity [57]. These capabilities have made AFM a widely used technique for characterizing brain tissue mechanics across tissue, cell, and even molecular scales [58, 59]. For example, Morr, et al. [60] used AFM to assess the microscopic mechanical properties of murine hippocampal subregions, and their results revealed that areas with high neurogenic activity exhibited nearly 40% lower stiffness than less active regions. Similarly, AFM measurements by Urbanski, et al. [61] on demyelinated mice and human brain tissue demonstrated that acute demyelination reduces stiffness, whereas chronic demyelination leads to increased stiffness.

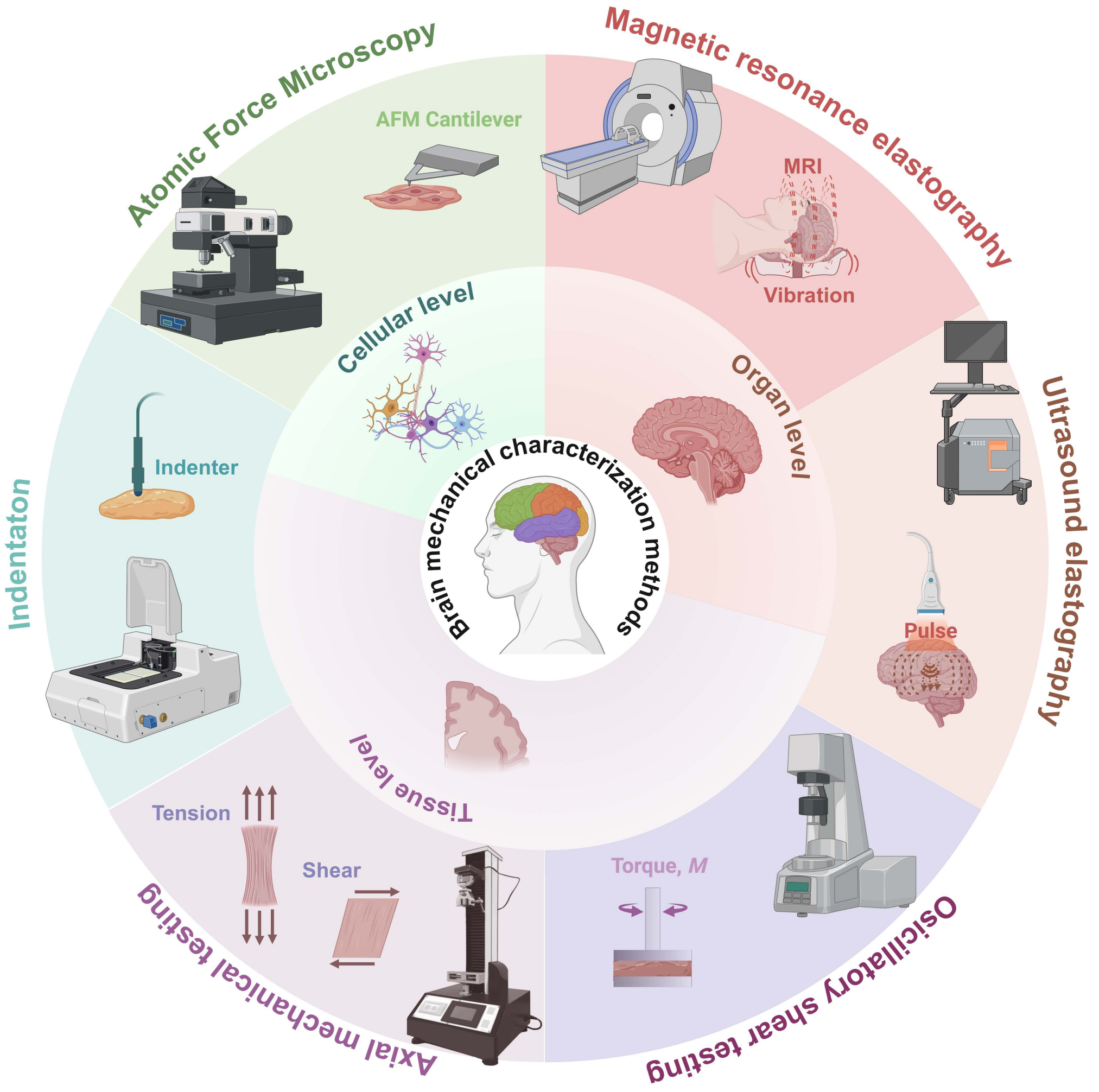

**Figure 1. Common mechanical characterization techniques for brain tissue**. Each method applies to distinct spatial scales. Figure created with BioRender.com.

Beyond mechanical characterization, AFM has been instrumental in elucidating pathogenic mechanisms of brain disorders from biomechanical perspectives [57, 62, 63]. De, et al. [64] conducted a comparative analysis using AFM to analyze amyloid-β protein aggregates in human cerebrospinal fluid, their observations support a correlation between the aggregate morphology and progression of AD. Lobanova, et al. [56] harnessed AFM's high-resolution capability to examine the size distribution of α-synuclein aggregates in cerebrospinal fluid and serum from Parkinson's disease (PD) patients, identifying a greater proportion of larger protein aggregates (exceeding 150 nm) in affected individuals. Additionally, Bahwini, et al. [55] employed AFM to compare the mechanical properties of cancerous and normal brain cells, observing significantly lower Young's modulus in cancer cells.

AFM can detect small variations in mechanical properties, making it a versatile tool for assessing tissue mechanical heterogeneity [65]. For instance, Elkin, et al. [66] used AFM to individually measure the elastic modulus of five subregions within the rat hippocampus and found significant regional variations in terms of the modulus value, highlighting the inherent mechanical heterogeneity of hippocampus tissue. Beyond elastic properties, AFM's versatility also allows it to characterize viscoelastic behavior. In addition to quasistatic indentation, AFM is well-suited for dynamic testing, including stress relaxation, strain creep, and oscillatory loading tests [67]. Due to its requirement for direct contact, AFM is primarily operated in *ex vivo* settings, such as on brain slices or isolated cells. A notable advancement came from Thompson, et al. [68], who developed time-lapse *in vivo* AFM (tiv-AFM) to measure changes in brain stiffness over time within a live embryo. Using this approach, they observed the stiffness-gradient-driven neuronal migration during early embryonic development in frogs. While AFM is effective for measuring brain mechanical properties, the testing outcome is influenced by the choice of tip geometry. Commonly used shapes include pyramidal, conical, and spherical indenters, each requiring different mechanical models for data interpretation. The Hertz model is typically applied for spherical indenters, the Sneddon model for conical indenters, and an extended Sneddon model for pyramidal indenters [27]. However, these models assume small deformation, and their accuracy may degrade when indentation exceeds these prerequisite limits. To address this limitation, alternative approaches such as the hyperelastic material model like the Ogden model [69] or the parameter reverse engineer approach [70] could be endeavored to improve parameter characterization accuracy. Additionally, as the tissue samples are excited from their native biological

environment—despite being preserved in supporting medium such as artificial cerebrospinal fluid—the measured properties may deviate from their in vivo state [27]. This limitation is also encountered in other invasive testing techniques to be introduced later.

**2.2. Indentation: scalable and versatile measurement of brain tissue mechanics**

IND is a frequently used technique for characterizing the mechanical properties of brain tissue. Similar to AFM, IND also works by measuring the contact behaviors between an indenter tip and the tissue surface. From a broader perspective, AFM can be treated as a scaled-down version of the indentation method [54]. As exemplified in Figure 2, a custom IND system is constructed by two main components: a loading cell equipped with a probe that applies the indentation force, and a displacement sensor that records the resulting deformation. With minimal sample preparation requirements, IND offers flexibility in measuring tissue mechanical properties at scales ranging from the microscale to the macroscale simply by adjusting the indenter size [71]. IND is frequently employed to assess the brain's regional stiffness within the elastic regime [28, 72, 73]. For instance, Weickenmeier, et al. [38] conducted 116 IND tests on bovine brain using a 1.5 mm diameter flat punch indenter, revealing that white matter is nearly twice as stiff as gray matter, with stiffness values of 1.330 ± 0.630 kPa and 0.680 ± 0.200 kPa, respectively. Expanding on this, Weickenmeier, et al. [74] used the same approach to measure stiffness in demyelination brain tissues. Combined with histological characterization, their findings indicated a positive correlation between white matter stiffness and myelin content, suggesting that brain tissue stiffness could serve as an effective biomarker for multiple sclerosis and other demyelinating brain disorders. More recently, Bailly, et al. [75] applied a smaller flat punch indenter (0.5 mm diameter) to characterize the elastic modulus of various spinal cord subregions. Their study identified significant heterogeneity in the elastic modulus of gray matter regions, while white matter regions exhibited more uniform stiffness.

In addition to healthy brain tissue, IND has been widely used to assess mechanical abnormalities in diseased brain tissue, including conditions such as brain tumors [76, 77], epilepsy [19], and AD [78]. Notably, Qian, et al. [79] took a step further in exploring the effects of electric fields—commonly introduced during brain disorder treatments—on brain mechanical properties. Using a flat punch indenter (8mm diameter), they conducted IDN tests on porcine brain tissue exposed to current electric field ranging from 0 to 50 V. Their results indicated that brain tissue softens and responds more rapidly at higher electric field intensities, contributing to potential refinements in therapeutic protocols. Beyond the elasticity measurements, IND has been employed to characterize the viscoelastic properties of brain tissue through stress relaxation and oscillatory loading tests [80-83]. Due to its simple tissue preparation and operational flexibility, IND can be performed not only *in vitro* but also *in situ* and even *in vivo* [84]. Prevost, et al. [85] conducted IND tests using a 12.65 mm diameter hemispherical indenter on the frontal and parietal lobes of living and deceased porcine brains after craniotomy, as well as on excited specimens, to measure various brain mechanical properties *in vivo*, *in situ*, and *in vitro*, respectively. Through testing, they found a significantly stiffer indentation response *in situ* than *in vivo*, implying a post-mortem stiffening effect [86]. In contrast, indentation responses *in vitro* exhibit greater compliance compared to *in situ* measurements.

Analogous to AFM, IND results are reliant on the choice of indenter tip geometry. Various shapes, including cylindrical [23], conical [84], spherical [83], rectangular [87], and square [73] indenter, have been used in existing studies. Notably, Budday, et al. [23] suggested a circular flat punch to minimize adhesion effects by maintaining a constant contact area between the indenter and tested samples, while Feng, et al. [88] recommending a rectangular indenter due to its asymmetric nature, which is beneficial for characterizing anisotropic mechanical properties. Additionally, indenter tip size also affects the testing outcomes. Budday, et al. [23] compared punch indenter with varying diameters ranging from 0.75 mm to 1.5 mm when measuring the bovine brain properties, finding that the elastic modulus decreased as the punch diameter increased. A similar trend was observed by Li, et al. [89], who reported that larger indenters significantly reduced the storage and loss stiffness of porcine brain tissue. Moreover, excessive indentation depth can introduce biases due to boundary effects from the substrates [90]. To minimize these effects and ensure accurate measurements, it is suggested that tissue thickness be at least 3-5 times greater than the indenter diameter [91] and that indentation depth not exceed 10% of the tissue's thickness [92]. Also, the accuracy and reliability of IND-based brain mechanical characterization are influenced by factors such as assumptions of incompressibility, isotropy, and frictionless contact, as well as the choice of mechanical models [71]. To improve these assessments, reverse engineering approaches using finite element method (FEM) can be employed to systematically evaluate and mitigate potential errors, ensuring a more precise representation of brain tissue mechanics [87, 93, 94].

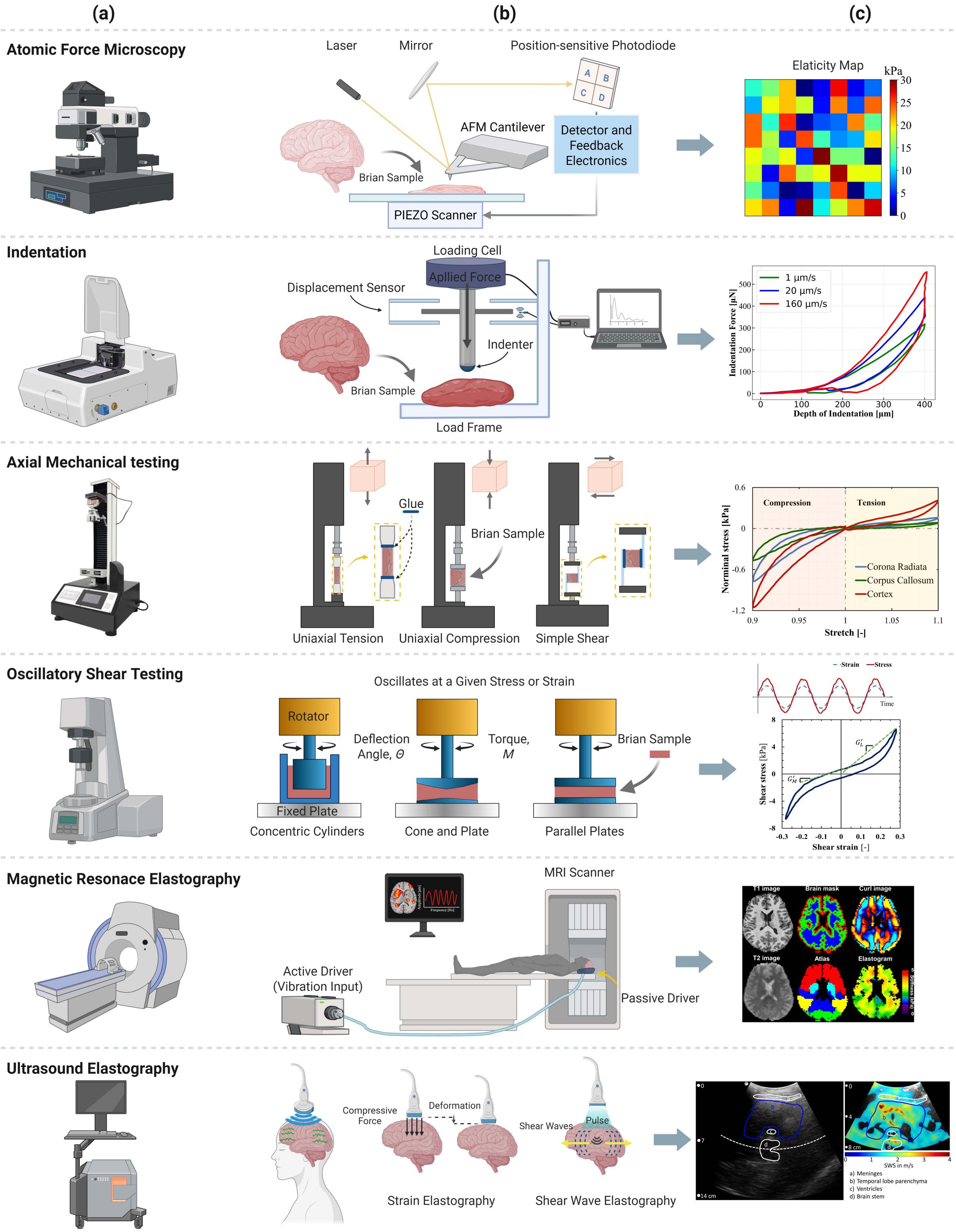

**Figure 2. Schematic representation showing the fundamental working principles of various testing techniques**. (a) Conceptual diagrams of testing apparatus; (b) Overview of the working mechanisms and special treatments applied during testing; (c) Key outcomes from each testing technique [95] Reproduced with permission. [96] Reproduced with permission. Figure created with BioRender.com.

## 2.3. Axial mechanical testing: mode-specific evaluation of brain tissue mechanics

During IND testing, the applied indentation force leads to nonuniform deformation within the brain tissue. The area directly beneath the indenter undergoes compression, while the surrounding tissue experiences tensile and shear forces to preserve structural integrity [73]. This interaction produces a complex stress state that blends brain tissue's response across different loading modes such as tension-compression asymmetry [30, 37]. An alternative approach, AMT—a well-established method in mechanical research—has been widely employed to obtain the mechanical properties of brain tissue under individual loading modes, including uniaxial tension, uniaxial compression, simple shear, and pure shear [45, 97]. AMT is performed by controlling the specimen holder to apply specific forms of displacement, thus the tissue requires to be securely attached to the holder during testing. Due to the fragile nature of brain tissue, glue adhesives are commonly adopted instead of traditional clamps [21]. In some cases, sandpaper is added to increase the adhesive surface area and ensure a more stable fixation [30]. Unlike IND testing, AMT engages the entire tissue in the loading process, resulting in nearly uniform force distribution across each cross-section. This uniformity allows for direct recording of stress-strain relationships, which significantly facilitates post-processing and ensures accurate characterizations of various mechanical properties, such as elastic modulus, yield strength, and failure strength [54]. Moreover, the consistent deformation patterns enable the investigation of brain mechanical anisotropy, which potentially arises from axonal fiber alignment or other structural factors [30, 41, 98]. This capability enriches the exploration of bridging the brain structural organization to its functions from a biomechanical perspective. However, to achieve these advantages, the tissue needs to be prepared in regular shapes, such as hexahedral or cylindrical forms with consistent cross sections, which inevitably complicates the preparation process. Additionally, this requirement makes it challenging to characterize regional material properties, particularly in small structures such as the hippocampus.

Extensive research has used AMT to characterize the mechanical properties of brain tissue under various loading conditions. For example, Miller, et al. [22] first conducted tension testing on swine brain tissue under finite deformation (< 20% strain) at two strain rates (0.64 and 0.0064 $s^{-1}$). Their findings demonstrated that brain tissue exhibits strain-rate dependence, stiffening with increased loading rates. Comparing these results with their previous compression test, they observed significantly softer behavior in tension and proposed an Ogden hyperelastic model to describe this asymmetry. A similar conclusion was derived by Rashid and his colleagues, who performed a series of tensile tests on porcine brain tissue up to 30% strain at higher strain rates ranging from 30 $s^{-1}$ to 90 $s^{-1}$ [99-101]. Notably, while Franceschini, et al. [24] recorded tensile failure in human brain tissue at about 90% strain, interior damage may initiate at much smaller strain (around 18% ) [21]. This suggests that careful consideration of strain limits is essential when conducting tensile tests on brain tissue. Additionally, to minimize boundary effects caused by adhesion during testing, sampling preparation should ensure a suitable aspect ratio. For cylindrical samples, the diameter-to-thickness ratio should ideally not exceed 1 [99], while for hexahedral samples, an equivalent dimensional balance should be considered. Ensuring an appropriate sample size helps prevent artifactual anisotropy caused by dimensional effect in mechanical characterization [30].

Although compression follows the opposite loading trend of tension, it is more versatile for measuring brain mechanical properties [45]. Given the biphasic nature of brain tissue (fluid vs solid about 4:1), compression test typically involves two modes: confined and unconfined compression. In confined compression, the fluid remains largely trapped within the solid matrix and contributes to tissue stiffness. In contrast, unconfined compression allows fluid to escape during the test, leaving the solid matrix to sustain the primary load [21]. Cheng, et al. [102] conducted unconfined compression tests on calf brain, revealing that the rheological response of white matter is primarily governed by the viscoelastic properties of the solid phase. A similar effect was observed by Su, et al. [103], who performed unconfined compression on the porcine brain and noted that this influence becomes more pronounced at low strain rates. In contrast, Haslach, et al. [104] performed confined compression test on rat brain tissue to isolate the contributions of the solid and fluid phases to brain mechanics. By forcing extracellular fluid to flow in the direction of deformation, they observed peak stress at about 11% strain, indicating that extracellular fluid plays a key role in load resistance until tissue damage permits pathological fluid flow. Their findings were further supported by magnetic resonance imaging, which revealed significant changes in tissue microstructure during confined compression. In confined compression, the tissue is radially constrained within a rigid, impermeable chamber to prevent outward movement. This requires bonding the tissue to the specimen holder, typically using surgical glue [104]. In unconfined compression, however, the tissue expands freely in the lateral direction without restraint. To facilitate this movement, lubricants like silicone grease are usually applied between the tissue and the holder to allow finite slippage [105]. For accurate modeling, Rashid, et al. [106] recommended dynamic friction coefficient values of 0.09 and 0.18 for strain rates of 1 $s^{-1}$ and 30 $s^{-1}$, respectively.

The simple shear test stands out for brain mechanical measurements due to its ability to better replicate physiological deformation compared to tension and compression. Shear tests distribute stress more evenly, thus reducing unwanted premature failure during testing. Rashid, et al. [107] conducted a simple shear test on porcine brain tissue at various loading rates and observed homogeneous deformation, which is further validated by the independence check of shear stress magnitude from specimen thickness. Similarly, Destrade, et al. [108] performed quasistatic simple shear tests on porcine brain tissues and compared their deformation behaviors to silicone gels. Their results indicated that brain tissue behaves as an extremely soft solid under shear force (at least 30 times softer than a silicone gel). Moreover, they identified a significant positive Poynting effect, meaning the brain tissue tends to "spread apart" perpendicular to the shear plane, generating compressive normal stress. This phenomenon was also observed in the torsional measurement of brain tissue [109]. To accurately capture this effect, Destrade, et al. [108] successfully modeled it using a two-term Mooney-Rivlin hyperelastic model. Additionally, Kuhl, et al. [110] suggested incorporating a second invariant of the deformation gradient into the strain energy function to improve the representation of this behavior.

Beyond pure uniaxial mechanical testing, many studies have employed multiple loading modes to better capture brain tissue mechanics. These include combinations of tension and compression [24, 93, 111], compression and shear [37, 42, 112, 113], or all three loading modes [26, 30, 98, 114]. These "multi-modal" tests provide a more comprehensive characterization of brain mechanical properties and greatly enhance the generalizability of calibrated models [21, 115, 116]. Additionally, multiaxial testing has been used to assess brain tissue anisotropy, though less frequently than in artery or skin studies. One of the few investigations was carried out by Labus and his colleagues, who performed biaxial tensile tests on Ovine brain tissue [117, 118]. In their studies, corona radiata and corpus callosum were extracted from white matter and subjected to biaxial tension to examine the role of axonal structure in brain mechanics. Using histology and transmission electron microscopy, they found a positive correlation between mechanical anisotropy and axon volume fraction. Furthermore, their findings suggest that combining both biaxial and uniaxial tests can significantly improve the accuracy of model predictions.

**Table 1. Comparison of techniques in measuring mechanical properties of brain tissue. Testing methods:** AFM: atomic force microscopy, IND: indentation, AMT: axial mechanical test, OST: oscillatory shear test, MRE: magnetic resonance elastography, USE: ultrasound elastography; **Testing modes:** QSIT: quasi-static indentation test, UT: uniaxial tension, UC: uniaxial compression, SS: simple shear, CT: creep test, SRT: stress relaxation test, CSRT: constant strain rate test, OLT: oscillatory loading test, CLT: cyclic loading test, FST: frequency sweep test, AST: amplitude sweep test, AT: adhesion test, MALT: multi-axial loading test, SWE: shear wave elastography, SE: strain elastography; **Measurable Mechanical Properties:** CL: cellular level, TL: tissue level, OL: organ level.

| **Testing Methods** | **Referenced Literatures & Brain Tissue** | **Experimental Settings** | | **Measurable Mechanical Properties** | **Sample Preparation Requirement** | **Accuracy-Sensitive Factors** |
|---|---|---|---|---|---|---|
| | | Testing Condition | Testing Modes | | | |
| AFM | • Canovic, et al. [58] mouse<br>• Bahwini, et al. [55] human<br>• Qing, et al. [119] mouse<br>• Ong, et al. [120] rat<br>• Iwashita, et al. [121] mouse<br>• Hall, et al. [122] rat<br>• Morr, et al. [60] murine<br>• Chuang, et al. [67] rat<br>• Najera, et al. [62] rat<br>• Runke, et al. [59] mouse | *in vitro*<br>*ex vivo* | QSIT, CT, SRT, OLT, AT, CLT | ▪ Young's modulus (CL)<br>▪ Shear modulus (CL)<br>▪ Viscoelasticity (CL)<br>▪ Hysteresis (CL)<br>▪ Surface adhesion (CL)<br>▪ Poisson's ratio (CL)<br>▪ Elasticity map (TL)<br>▪ Heterogeneity (TL) | ▪ Cut fresh tissue into thin sections for easy handling and indentation<br>▪ Ensure a flat sample surface<br>▪ Keep sample hydrated to prevent dehydration and mechanical variation<br>▪ Control temperature<br>▪ Immobilize tissue to minimize movement during measurement | ▪ Probe tip geometry<br>▪ Surface roughness<br>▪ Substrate effects<br>▪ Indentation depth<br>▪ Tip/sample adhesion<br>▪ Mechanical model assumptions |
| IND | • Budday, et al. [23] bovine<br>• Stewart, et al. [76] human<br>• Weickenmeier, et al. [74] bovine<br>• MacManus, et al. [123] human<br>• Qian, et al. [79] porcine<br>• Menichetti, et al. [28] human<br>• Sundaresh, et al. [82] porcine<br>• Pan, et al. [19] human<br>• Basilio, et al. [94] human<br>• Bailly, et al. [75] porcine | *ex vivo,*<br>*in vitro,*<br>*in situ,*<br>*in vivo* | QSIT, CT, SRT, OLT, CLT, MALT | ▪ Young's modulus (TL)<br>▪ Shear modulus (TL)<br>▪ Viscoelasticity (TL)<br>▪ Hysteresis (TL)<br>▪ Hardness (TL)<br>▪ Poisson's ratio (TL)<br>▪ Heterogeneity (TL) | ▪ Select bulk tissue samples with proper thickness to mitigate substrate effects<br>▪ Ensure a flat and smooth tissue surface<br>▪ Maintain tissue hydration<br>▪ Control temperature<br>▪ Secure the sample to prevent movement during measurement | ▪ Indenter shape and size<br>▪ Indentation depth<br>▪ Surface roughness<br>▪ Preconditioning effects<br>▪ Mechanical model assumptions<br>▪ Boundary effects and constraints |
| AMT | • Miller, et al. [22] porcine<br>• Hrapko, et al. [113] porcine<br>• Jin, et al. [98] human<br>• Rashid, et al. [101] porcine<br>• Destrade, et al. [108] porcine<br>• Budday, et al. [30] human<br>• Budday, et al. [124] human<br>• Budday, et al. [125] human<br>• Balbi, et al. [109] porcine<br>• Hosseini-Farid, et al. [126] porcine<br>• Hosseini-Farid, et al. [127] porcine<br>• Eskandari, et al. [111] bovine<br>• Boiczyk, et al. [37] porcine<br>• Su, et al. [103] Porcine | *in vitro,*<br>*ex vivo* | UT, UC, SS, CT, SRT, CLT, CSRT, MALT | ▪ Young's modulus (TL)<br>▪ Shear modulus (TL)<br>▪ Tissue strength (TL)<br>▪ Viscoelasticity (TL)<br>▪ Hyperelasticity (TL)<br>▪ Hysteresis (TL)<br>▪ Fatigue resistance (TL)<br>▪ Poisson's ratio (TL)<br>▪ Heterogeneity (TL)<br>▪ Anisotropy (TL) | ▪ Select uniform tissue samples to ensure consistent mechanical properties<br>▪ Cut samples into standardized sizes, ensuring parallel and smooth loading surfaces<br>▪ Control temperature<br>▪ Fix the sample securely to prevent slippage or uneven loading<br>▪ Ensure alignment of the sample along the loading axis<br>▪ Measure initial dimensions (length and cross-sectional area) to ensure accurate stress-strain calculations | ▪ Preconditioning effects<br>▪ Clamping artifacts<br>▪ Sample slippage<br>▪ Hydration loss<br>▪ Sample misalignment<br>▪ Mechanical model assumptions |

(*Continued*)

**Table 1. Comparison of techniques in measuring mechanical properties of brain tissue. Testing methods:** AFM: atomic force microscopy, IND: indentation, AMT: axial mechanical test, OST: oscillatory shear test, MRE: magnetic resonance elastography, USE: ultrasound elastography; **Testing modes:** QSIT: quasi-static indentation test, UT: uniaxial tension, UC: uniaxial compression, SS: simple shear, CT: creep test, SRT: stress relaxation test, CSRT: constant strain rate test, OLT: oscillatory loading test, CLT: cyclic loading test, FST: frequency sweep test, AST: amplitude sweep test, AT: adhesion test, MALT: multi-axial loading test, SWE: shear wave elastography, SE: strain elastography; **Measurable Mechanical Properties:** CL: cellular level, TL: tissue level, OL: organ level.

| Testing Methods | Referenced Literatures & Brain Tissue | Experimental Settings: Testing Condition | Experimental Settings: Testing Modes | Measurable Mechanical Properties | Sample Preparation Requirement | Accuracy-Sensitive Factors |
|---|---|---|---|---|---|---|
| OST | • Shuck, et al. [128] human<br>• Arbogast, et al. [129] porcine<br>• Darvish, et al. [130] bovine<br>• Nicolle, et al. [112] porcine & human<br>• Hrapko, et al. [131] porcine<br>• Garo, et al. [92] porcine<br>• Chatelin, et al. [132] human<br>• Canovic, et al. [58] mouse<br>• Li, et al. [133] porcine<br>• Boudjema, et al. [29] lamb & cow<br>• Qing, et al. [119] mouse<br>• Xue, et al. [134] rat | *in vitro, ex vivo* | OLT, FST, AST, CT, SRT | ▪ Viscoelasticity (TL)<br>▪ Storage modulus (TL)<br>▪ Loss modulus (TL)<br>▪ Damping factor (TL)<br>▪ Yield strain (TL) | ▪ Select uniform tissue samples to ensure consistent mechanical properties.<br>▪ Trim samples into a well-defined shape for uniform shear strain distribution<br>▪ Ensure consistent sample thickness to minimize boundary effects<br>▪ Maintain physiological hydration<br>▪ Control temperature<br>▪ Fix the sample securely to prevent slippage or uneven loading<br>▪ Measure the sample's initial dimensions (diameter, thickness) for accurate shear stress calculations | ▪ Preconditioning effects<br>▪ Non-uniform tissue thickness<br>▪ Sample slippage<br>▪ Plate surface properties<br>▪ Sample off-center placement<br>▪ Strain amplitude selection<br>▪ Mechanical model assumptions |
| MRE | • Murphy, et al. [95] human<br>• Johnson, et al. [135] human<br>• Bigot, et al. [136] rodent<br>• Huang, et al. [137] human<br>• Yeung, et al. [138] human<br>• Smith, et al. [139] human<br>• McIlvain, et al. [140] human<br>• Bergs, et al. [43] human | *in vivo, in situ* | OLT, FST | ▪ Shear modulus (OL)<br>▪ Bulk modulus (OL)<br>▪ Viscoelasticity (OL)<br>▪ Stiffness map (OL)<br>▪ Heterogeneity (OL)<br>▪ Anisotropy (OL) | ▪ No samples preparation required | ▪ MRI scanner capability<br>▪ Compressive wave interference<br>▪ Shear wave frequency<br>▪ Imaging resolution<br>▪ Motion artifacts |
| USE | • Xu, et al. [141] human<br>• Liu, et al. [142] porcine<br>• Liu, et al. [143] human<br>• Lay, et al. [144] mouse<br>• Garcés Iñigo, et al. [145] human<br>• Klemmer Chandía, et al. [146] human | *in vivo, in situ, ex vivo* | SWE, SE | ▪ Shear modulus (TL&OL)<br>▪ Viscoelasticity (TL&OL)<br>▪ Viscosity (TL&OL)<br>▪ Stiffness map (TL&TL)<br>▪ Heterogeneity (TL&OL) | ▪ No samples preparation required for *in vivo* or *in situ* testing. If used *ex vivo*, critical requirements are as follows:<br>▪ Keep samples hydrated during testing<br>▪ Apply ultrasound coupling gel evenly to ensure proper wave transmission<br>▪ Ensure flat and even contact between the transducer and tissue | ▪ Shear wave frequency<br>▪ Transducer positioning<br>▪ Skull Attenuation effects<br>▪ Boundary reflection |

## 2.4. Oscillatory shear testing: frequency-dependent insights into brain tissue mechanics

Brain tissue primarily undergoes shear deformations under physiological and pathological conditions, such as during impact trauma. This makes shear testing a more relevant method for assessing brain mechanics compared to tension and compression. In addition to the quasistatic shear test, OST has been widely used to characterize the viscoelastic behaviors of brain tissue. As illustrated in Figure 2, OST operates by using a rotating component to apply oscillatory motion, which is transmitted to the adhesive tissue sample [132]. Unlike shear tests in AMT, where tissue is subjected to continuous monotonic shear at constant strain rates, OST oscillates tissue back and forth by applying cyclic shear, typically in the form of sinusoidal strain or stress [132, 133]. Depending on how the sinusoidal input is configured, OST can be categorized into amplitude sweep tests (AST) and frequency sweep tests (FST). In AST, the maximal shear is systematically varied over a predefined range at a fixed frequency. In contrast, FST keeps the shear amplitude constant while varying the frequency of oscillation [134]. These tests enable direct quantification of the brain tissue's viscoelastic properties, including the storage modulus (elastic response) and loss modulus (viscous response). The ratio between them, known as the loss tangent, reflects the balance between elastic and viscous behavior, indicating whether the tissue behaves more like a solid or a fluid [42].

Shuck, et al. [128] were among the first to systematically conduct FST on human brain tissue, applying shear strain at 3.5% across a frequency range of up to 350 Hz. Using a four-parameter linear viscoelastic model to calibrate their results, they derived a set of frequency-dependent storage and loss moduli and observed a frequency-stiffening trend for both. More recently, Xue, et al. [134] employed AST to investigate the age-dependent viscoelastic properties in rat brain tissues across developmental stages, from postnatal day 4 to 4 months of age. They first conducted AST at 0.16 Hz with strain amplitude sweeping from 0.01% to 100% to determine the linear viscoelastic range. A 1% shear strain threshold was identified as the upper limit of this range and subsequently used for FST over a frequency span from 0.016 Hz to 19.1 Hz. Results showed that both storage and loss moduli increased with age. However, the ratio of the two moduli remained constant at low frequencies (<1.6 Hz) and began to decline in the mid-frequency range (1.6–16 Hz). In a related study, Qing, et al. [119] applied a similar approach to assess frequency-dependent shear moduli in both healthy and autism spectrum disorders (ASD) rat brain tissues.

AST was first performed at 0.16 Hz and 1.6 Hz to determine a linear viscoelastic limit between 1–3% shear strain. FST was then carried out at 1% strain across frequencies from 0.016 Hz to 6.4 Hz. Their findings revealed no significant differences in the loss and storage moduli of ASD brain tissue, in spite of significant changes observed in cellular organization.

Most OSTs on brain tissue have been conducted at small strain amplitudes to minimize damage to this fragile material. These conditions correspond to the linear viscoelastic range, where linear models such as the Kelvin-Voigt model can be effectively applied to characterize material properties [58, 92, 112, 129, 133]. However, brain tissue exhibits inherent nonlinear behavior, suggesting that its viscoelastic response under large deformation may differ significantly from that observed under small deformation [41, 125]. In scenarios such as automotive crashes, the brain can experience large shear deformations (often exceeding 10-20% strain) within milliseconds. Under such conditions, viscoelastic properties characterized using linear models may fail in accurately capturing brain behaviors [147]. To address this discrepancy, Darvish, et al. [130] conducted FST on bovine brain tissue across a frequency range of 0.5 to 200 Hz at a shear strain of up to 20%. A quasilinear viscoelastic model was employed to account for nonlinear behaviors at large strains. Their results exhibited a pronounced strain-hardening effect beyond 10% strain and identified a non-recoverable strain conditioning behavior in measured moduli, which may explain discrepancies reported across different studies. Analogously, Boudjema, et al. [29] performed larger-deformation shear tests on lamb and bovine brain tissue, measuring storage and loss moduli over a frequency range of 1 to 100 Hz and at shear strains up to 50%. Their findings also revealed clear strain-dependence behavior and a notable stiffening effect in both moduli. This observation, however, contrasts with traditional views that brain tissue softens at large deformations due to tissue damage or degradation [112, 131]. These contradictory findings may stem from differences in the viscoelastic models used to interpret the data, as Boudjema, et al. [29] did not explicitly report the constitutive models applied in their analysis. In addition to shear testing, oscillatory methods have also been applied in other loading modes, such as compression [39, 148], which further demonstrates the versatility of oscillatory testing in characterizing brain tissue mechanics.

### 2.5. Magnetic resonance elastography: *in vivo* measurement of brain tissue mechanical properties

MRE is a phase-contrast MRI technique that allows for the noninvasive assessment of brain tissue mechanics *in vivo* [20, 44, 149]. The process typically involves three components: a mechanical actuator that generates a shear wave within the brain, an MRI scanner that captures tissue deformations, and a post-processing system that estimates mechanical properties based on imaging data [135]. As illustrated in Figure 2, shear waves are commonly produced using a passive mechanical actuator, such as a soft pillow, placed. During testing, the pillow is placed beneath the head and continuously vibrates at predefined frequencies controlled by an external active pneumatic driver [137, 150]. Alternative actuator designs include a head cradle [151], bit-bar [136], and customized piezoelectric soft actuators [152]. In contrast to these external actuators, a pilot study by Weaver, et al. [153] proposed an intrinsic activation approach that uses natural cardiac pulsations. By capturing blood flow oscillations via MR angiography and converting them into harmonic deformations through Fourier transformation, this method skillfully circumvented the issue of wave attenuation or delay caused by the cranium and meninges [154]. As a result, it offers more consistent and reproducible measurements of brain mechanical properties. For external actuation, the actuation frequency is often controlled within a range of 10-100 Hz [155], typically around 50-60 Hz considering the balance between penetration and attenuation attributes of indued shear waves [156]. Intrinsic actuation, on the other hand, enables much lower frequencies (around 1Hz), which significantly benefits for probing deep brain regions. Because lower frequencies correspond to longer wavelengths that experience less attenuation, thereby improving signal penetration and reducing imaging noise [157]. While promising progress has also been made in imaging algorithms, such as motion encoding, phase image preprocessing, and material property reconstruction, these topics will not be elaborated here as they fall beyond the scope of this review. Readers interested in these technical advancements are referred to the comprehensive reviews by Hiscox, et al. [157] and Johnson, et al. [135].

In brain MRE, various mechanical properties, such as Young's modulus, storage and loss moduli, as well as bulk modulus, can be characterized based on the underlying material model assumptions, including elasticity [140, 158], viscoelasticity [159, 160], and poroelasticity [153, 156] (see Table 1). Thanks to its noninvasive nature and modeling versatility, MRE has been widely employed in both fundamental and clinical brain research. For example, Sack, et al. [161]

used MRE to investigate the effects of aging and gender on brain viscoelastic properties in a cohort of 55 healthy participants (23 females) aged 18 to 88 years. Their study for the first time revealed a notable decline in brain shear stiffness with age, approximately 0.8% per year. Interestingly, they also found that female brains were, on average, 9% stiffer than male brains, implying that women may be mechanically a decade "younger" than men. Similar age-related trends have been reported in other studies [138, 140, 151, 162]. With the incorporation of brain parcellation atlas, MRE has been further used in examining the regional mechanical difference across brain structures, from broad distinctions between gray and white matter [137, 151, 163, 164] to more detailed explorations of functionally diverse regions [160, 165-167]. Clinically, MRE holds great promise for facilitating the diagnosis of brain disorders [168]. It has been applied to characterize mechanical changes in conditions such as brain tumors [158, 169-171], neurodegenerative disease like AD [155, 172] and PD [173], brain injury [44], dementia [174], normal pressure hydrocephalus [175], and epilepsy [176]. In addition, advanced techniques such as multi-excitation MRE can generate and process shear waves from multiple directions. When combined with a nonlinear FEM-based inversion algorithm, this MRE approach enables the measurement of anisotropic mechanical properties, particularly in white matter [139, 177, 178].

While MRE is a powerful in measuring brain mechanical properties *in vivo*, several challenges remain that limit its accuracy and interpretability. A key limitation lies in its spatial resolution, which is often insufficient to capture fine-scale anatomical features and the heterogeneity of brain tissue, especially in small regions such as the hippocampus or deep gray matter structures [43]. To address this, the use of higher magnetic field strengths (e.g., 7T) to improve the signal-to-noise ratio [166, 167], optimized motion encoding gradients [179], and advanced imaging sequences like spin-echo echo-planar imaging [180] may enhance imaging spatial fidelity. Another major challenge is the inverse problem of MRE, namely deriving mechanical properties from the measured wave fields. This problem, however, is mathematically ill-posed and highly sensitive to noise, boundary conditions, and model assumptions, such as tissue homogeneity and linear viscoelasticity [157]. To tackle this issue, nonlinear inversion algorithms implemented using FEM may significantly enhance the physical rigor of the results [181]. Nonetheless, FEM-based approaches also substantially increase the computational time required for postprocessing [182]. Recent advances in deep learning-based methods offer a promising alternative by accelerating the inversion process while maintaining or even boosting effective accuracy [183, 184].

### 2.6. Ultrasound elastography: real-time assessment of brain tissue mechanics

Another popular elastography technique is USE, which takes advantage of ultrasound's stiffness-sensitivity to estimate the mechanical properties of soft tissues [185, 186]. Compared to MRE, USE is capable of providing real-time imaging of tissue mechanical properties using a cost-effective system with greater portability and accessibility, making it well-suited for bedside diagnostic applications. Because the skull impedes ultrasound transmission and hinders effective imaging of intracranial structures, USE has not been as widely used to assess the mechanical properties of brain tissue compared to other organs such as the liver, breast, and kidney [187]. Despite this challenge, significant efforts have been dedicated to advancing transcranial USE techniques to enable reliable and clinically relevant assessment of brain mechanics [188].

The technical details of USE for measuring brain mechanical properties have been documented in several studies [185, 187, 189]. Based on the operational principles, brain-focused USE techniques are generally divided into two categories: quasistatic strain elastography (SE) and dynamic shear wave elastography (SWE) [189], as shown in Figure 2. In SE, axial brain deformation, induced either by external compressive forces or internal physiological stimuli such as cardiac pulsation, is captured by ultrasound to estimate tissue mechanical properties [190]. This approach typically produces a 2D strain map derived from B-mode ultrasound images, offering relative comparisons of tissue stiffness rather than absolute mechanical values. For example, Kim, et al. [191] used SE to assess 21 healthy neonates between 28 and 40 gestational weeks. Using semi-quantitative color scale assessment, they compared the relative stiffness of various brain regions, including ventricle, periventricular white matter, caudate, subcortical, cortical gray matter, and subdural space. Their results showed notable stiffness variation across brain regions, with cortical gray matter being the stiffest region, and found a positive correlation between tissue stiffness and gestational age. Despite these insights, the qualitative nature of SE and its inability to provide absolute stiffness values restrict its broader applicability in scientific and clinical contexts.

In contrast to SE, SWE enables quantitative assessment of tissue mechanical properties such as shear modulus [192], viscoelasticity [96], and hyperelasticity [193] (see Table 1). Like

MRE, SWE relies on the generation of shear waves within brain tissue to infer material properties. These waves can be generated either mechanically or through acoustic radiation force (ARF). One mechanical approach is transtemporal time-harmonic elastography (THE), which uses externally induced harmonic vibrations along with ultrasound-based motion tracking to image brain properties [194]. Klemmer Chandía, et al. [146] conducted a comparative study using both multifrequency MRE (20-35 Hz) and THE (27-56 Hz). Their findings indicated that THE can provide brain stiffness measurements consistent with MRE. Moreover, they suggested an optimal THE measurement depth of 40-60 mm, balancing ultrasound attenuation and near-field effects. The ARF-based method uses high-intensity focused ultrasound (in the MHz range) to create localized tissue displacement in the axial direction. These displacements further generate both compressive (impulse) waves and lateral shear waves that propagate through the brain tissue. Compressive waves are used in acoustic radiation force impulse imaging (ARFI), which infers tissue stiffness by measuring axial displacement using techniques such as virtual touch quantification (VTQ) [195]. In contrast, lateral shear waves are analyzed to quantify stiffness based on the assumption that shear wave speed correlations with tissue stiffness [189].

USE is generally easier to operate and requires less specialized expertise than MRE. Its measurement is also less sensitive to variations in signal-to-noise ratio, making USE a more practical and accessible option for characterizing brain tissue mechanical properties in pediatric populations [144, 145, 189, 191, 195]. Additionally, USE has shown promise in diagnosing and monitoring a range of brain conditions, including brain tumors [196, 197], ischemic stroke [141], TBI [198, 199], and hydrocephalus [200], due to its sensitivity to stiffness changes associated with pathological processes. The versatility of USE also extends to *ex vivo* applications, particularly on animal brains. Liu, et al. [143] performed both *in vivo* and *ex vivo* SWE measurement on rabbit brains. *In vivo* tests were conducted after removing the skin and skull over the frontal lobe, while *ex vivo* measurements were taken on dissected brain tissue immersed in artificial cerebrospinal fluid at body temperature. Their results exhibited an average 47% increase in shear modulus *in vivo* compared to *ex vivo* measurements, although this difference became negligible when *ex vivo* tests were taken within 60 minutes of tissue extraction. Despite its advantages, USE faces several limitations. In addition to skull-induced acoustic impedance and scattering, ultrasound experiences significant attenuation within brain tissue, limiting its ability to probe deep brain structures. Additionally, ultrasound measurement is also highly operator-dependent, which can affect measurement accuracy and reproducibility. For example, Blackwell [187] demonstrated that stiffness measurements in bovine brains were highly sensitive to the angle of the transducer application. Moreover, USE has limited capability in accurately characterizing the heterogeneity of brain tissue, as complex shear waves propagation and potential mode conversions can distort elastographic measurements and complicate data interpretation [201].

## 3. Discussions of Human Brain Mechanical Testing

Accurate mechanical characterization of human brain tissue is essential for understanding the complex biomechanical behaviors of the brain under both physiological and pathological conditions. Such testing, however, is inherently challenging due to ethical considerations and the limited availability of suitable human samples [21]. These constraints have led researchers to explore the use of animal brain tissue as an experimental surrogate, with species such as porcine and bovine commonly employed based on reported similarities in gross anatomy or mechanical response [38, 85, 101, 127]. Despite their widespread use, there still remains no clear consensus regarding the appropriateness of animal models for replicating human brain tissue mechanics. For example, MacManus, et al. [33] advocated for the use of pig and rat brains based on a comparative analysis of dynamic mechanical properties among mouse, rat, pig, and human brains. In contrast, Prange, et al. [41] found that human brain tissue exhibited significantly greater stiffness than porcine brain, with an average 29% higher shear modulus. Variabilities in genetic backgrounds, anatomical structures, and cellular compositions between animals and humans can further contribute to substantial deviations in mechanical behaviors [40]. An illustrative example is found in the contrasting regional stiffness trends reported by Budday and her colleagues. In their bovine brain study, white matter (average modulus of $1.895 \pm 0.592$ kPa) was found to be approximately 39 % stiffer than gray matter (average modulus of $1.389 \pm 0.289$ kPa) [23]. However, their subsequent study on human brain tissue revealed the opposite pattern, with gray matter regions (average modulus of 1.065 kPa) exhibiting nearly double the stiffness of white matter regions (average modulus of 0.505 kPa) [30]. In addition, animal brain models often fall short in fully capturing the complex, anisotropic, and region-specific mechanical characteristics intrinsic to the human brain [34, 42]. These limitations are particularly problematic in the context of computational modeling and constitutive model characterization [202, 203], where accurate mechanical data from human brain tissue are vital for developing realist computational models to simulate fundamental brain mechanisms [4, 5, 204], injury biomechanics [8], surgical interventions [18], and disease progression [205] with high fidelity.

Given the challenges pertinent to human brain tissue mechanical measurement, each related study represents a valuable contribution toward unraveling the complexity and inherent elegance of human brain biomechanics. In what follows, we reviewed the body of literature dedicated to mechanical testing of human brain tissue, aiming to provide a comprehensive and comparative overview of its mechanical characteristics. To structure our analysis, we categorized the testing methods into two main groups: invasive and noninvasive approaches. Relevant publications were identified through an extensive literature search using widely accessed academic databases, including Google Scholar, PubMed, and Web of Science. The keywords used for searching include "human brain", "mechanical testing", "mechanical properties", and "material characterization". Notably, the following review primarily focuses on invasive testing methods, as they directly probe the material response of brain tissue and offer more robust quantitative data for constitutive modeling and validation of computational simulations.

### 3.1. Invasive mechanical testing: direct, informative, and high-fidelity characterization

Through our literature search, we identified 34 peer-reviewed studies that employed invasive mechanical testing on human brain tissue, spanning from the earliest work by Fallenstein, et al. [206] to the recent advancements reported by Greiner, et al. [207]. The details of these studies are presented in Table 2, where we summarize key experimental parameters, including sampling regions; tissue freshness (indicated by post-mortem interval or durations after surgical resection); subject age distribution; specimen geometry; measurement techniques; loading modes performed in testing; environmental testing temperature; frequencies or strain rates chosen for dynamic test and predefined strain range. Additionally, we documented the reported materials properties and constitutive models used for data interpretation, as well as the parameter calibration methods, whether through conventional least squares fitting or inverse FEM-based indentation approaches. Based on this comprehensive dataset, we analyzed the distribution of key experimental attributes to find common trends across studies. We further discussed the primary factors that contribute to variabilities in reported properties, such as differences in specimen preparation, experimental protocols, and modeling assumptions. To facilitate a comparative overview, we also summarized the mechanical properties extracted from these studies, including various moduli (e.g., shear, storage, loss, and relaxation modulus) and representative stress-strain curves. Through these analyses, we aim to address the following two questions: (1) what factors contribute to the huge variability in reported mechanical properties of the human brain? (2) what insights can we infer from existing experimental efforts regarding the mechanical behaviors of human brain tissue?

Figure 3 illustrates a statistical summary of several key aspects observed across the reviewed studies, including testing regions (Figure 3a), testing methods (Figure 3b), fitted material models (Figure 3c), and primary loading modes (Figure 3d). As shown, the cortex (C) is the most frequently tested human brain region, likely due to its anatomical location as the outermost layer of the brain, which makes it more accessible for sample collection during surgical resections or post-mortem dissections. The corona radiata (CR) and corpus callosum (CC) are also commonly selected, owing to their well-defined structure as major white matter tracts—ideal candidates for investigating the anisotropic mechanical properties of the brain. Notably, the category labeled cerebrum (Cb) refers to samples containing both gray and white matter, typically used in earlier studies where tissues were undissected or only roughly separated without clear anatomical differentiation [206, 208-210]. Specimens classified as diseased brain (DB) primarily include brain tumors, though some studies have also examined the biomechanical implications of neurological conditions such as epilepsy [19, 211] and AD [212]. Among the various testing methods, AMT and IND stand out as frequently used methods due to their experimental versatility and ability to characterize localized mechanical response (see Sections 2.2 and 2.3). OST has been widely utilized to capture the frequency-dependent viscoelastic behavior of brain tissue, which is particularly relevant for injury biomechanics such as automotive impact modeling [42, 112, 128, 132, 206]. Only one study by Park, et al. [212] employed AFM to assess viscoelastic properties of autopsy brain tissue in the context of AD. In terms of mechanical parameters calibration, the human brain is commonly represented as viscoelastic or hyperelastic, reflecting its nonlinear and time-dependent response. Only a few studies have assumed linear elasticity, primarily to estimate basic elasticity properties such as Young's modulus [77, 213]. Additionally, the poroelasticity of the brain has been explored to account for its biphasic features and fluid-solid interactions [24, 42, 93, 214]. A variety of loading modes have been implemented in human brain tissue testing. Among them, stress relaxation tests are commonly used due to their suitability for capturing the viscoelastic response of brain tissue. In addition, compression and shear tests are more frequently performed compared to tension tests, likely because of the highly fragile and easily damaged nature of brain tissue, which poses significant challenges for tensile loading [97].

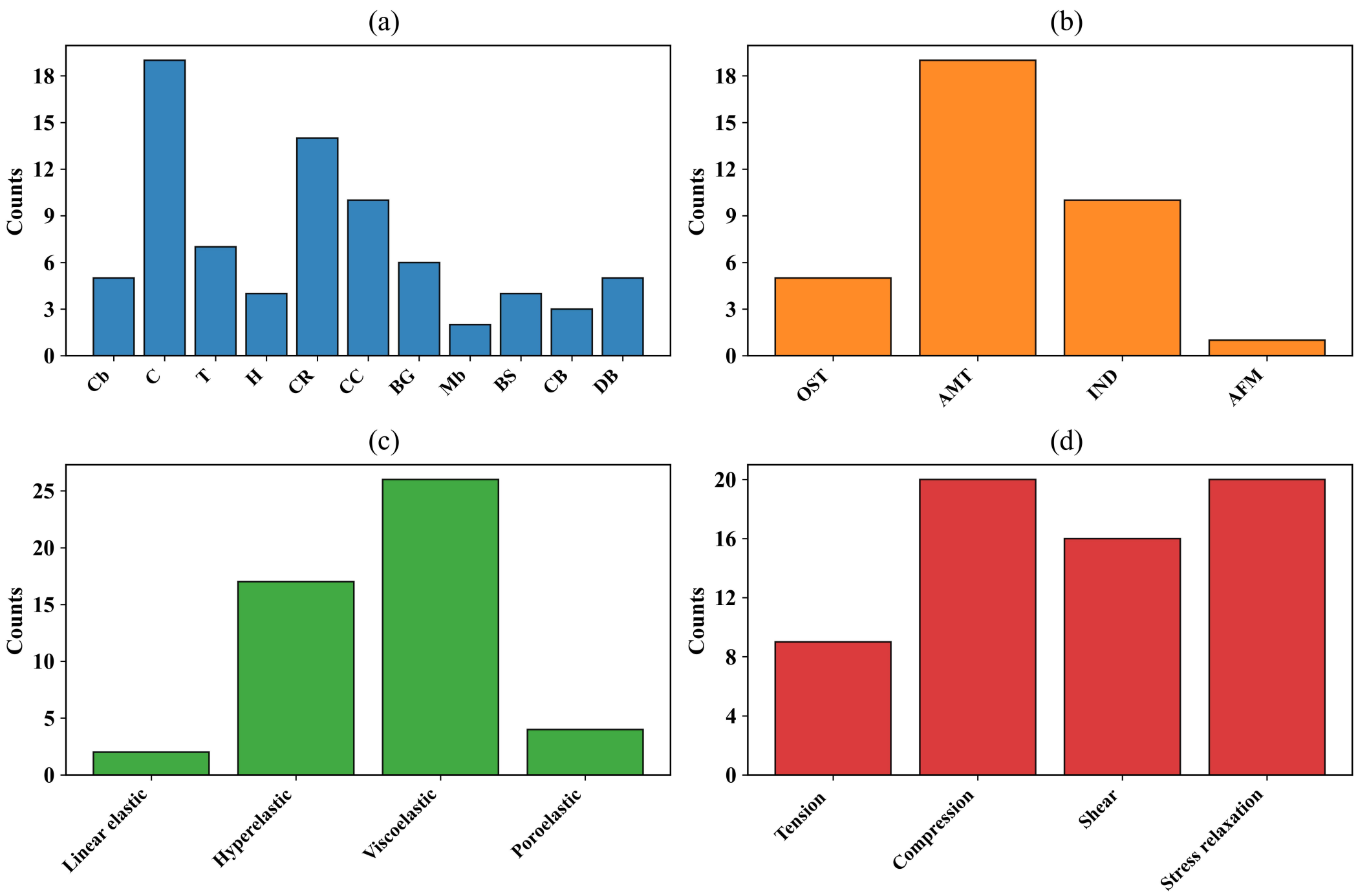

**Figure 3. Summary of publications on human brain tissue mechanical testing**. (a). Distribution of tested brain regions; (b). Distribution of testing methods used for measuring human brain tissue mechanical properties; (c). Model assumptions applied in characterizing brain tissue mechanical properties; (d). Loading modes performed in human brain tissue mechanical testing. Abbreviations in (a) and (b) follow those in the caption of Table 1.

Despite extensive efforts that have been made to characterize the mechanical properties of human brain tissue, the reported parameters show substantial variability, often spanning several orders of magnitude, and in some cases, even present contradictory findings across studies [71]. Figure 4 compares various shear-related moduli, including shear modulus, instantaneous modulus, relaxation modulus, and long-term modulus, between gray and white matter, based on existing literature focused on human brain tissue. For studies that investigated multiple brain regions, shear modulus values were averaged across gray and white matter regions to facilitate comparison. Additionally, for studies reporting only storage and loss moduli, the magnitude of the complex modulus was calculated to represent an equivalent shear modulus. Since Shuck, et al. [128] and Chatelin, et al. [132] reported frequency-dependent complex shear moduli, we present only the values measured at the lowest frequencies used in their studies—5 Hz and 0.1 Hz, respectively. Only studies that reported mechanical properties for both gray and white matter are included to enable direct inter-group comparisons. In this analysis, the relaxation modulus is defined as the difference between the instantaneous and long-term shear modulus, reflecting the gradual reduction in shear resistance due to viscous dissipation. As seen in figure4, reported values of shear modulus vary in multiple orders of magnitude, from tens of Pascals to several kilopascals, especially for the instantaneous and long-term components. Moreover, conflicting trends between gray and white matter properties are also evident. For example, Zhu, et al. [25] reported that gray matter is softer than white matter, with stiffness values of 3,100 Pa and 4,100 Pa, respectively (Figure 4a). This finding is supported by Pan, et al. [19], who found gray matter (653.36 ± 155.81 Pa) only to be slightly softer than white matter (684.58 ± 101.61Pa). In contrast, studies by Budday and her colleagues reported the opposite trend, indicating a significantly higher stiffness value for gray matter compared to white matter [26, 30, 114]. Similar inconsistencies in the relative stiffness of gray and white matter are also observed for the other shear moduli shown in Figures 4b-4d.

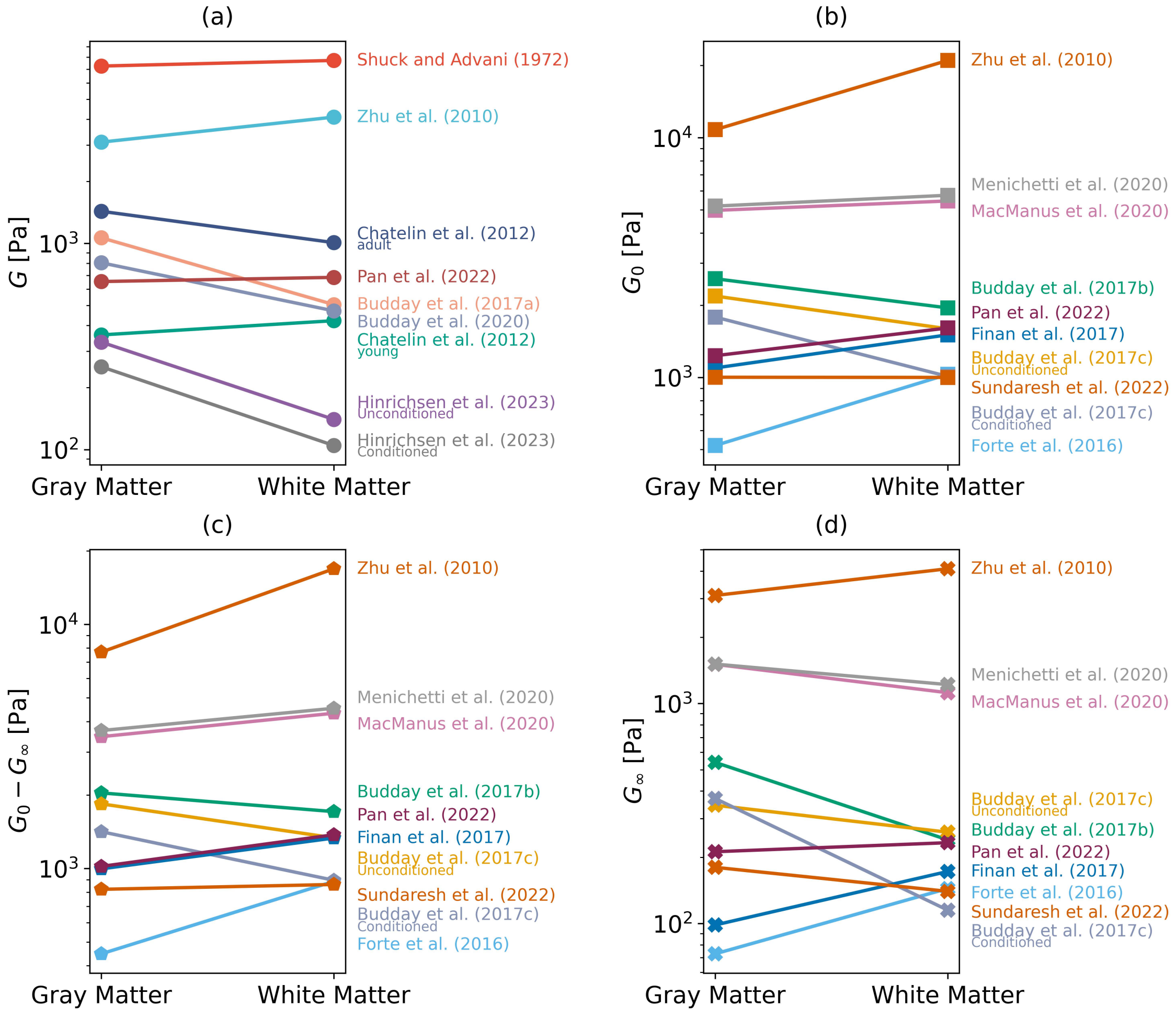

**Figure 4. Gray matter vs white matter in measured mechanical properties**. (a) Comparisons of shear modulus ($G$); (b-d). Comparisons of instantaneous shear modulus ($G_0$), relaxation shear modulus ($G_0 - G_\infty$), and long-term shear modulus ($G_\infty$). Only invasive studies are included, as collected in Table 2. For studies that measure multiple gray or white matter regions, modulus values are averaged across regions.

**Table. 2. Literature summary of human brain tissue testing**. **Regions**: Cb: cerebrum (white and gray), C: cortex, T: thalamus, H: Hippocampus, CR: corona radiata, CC: corpus callosum, BG: basal ganglia, Mb: midbrain, BS: brain stem, CB: cerebellum, WB: whole brain, BT: brain tumor; **PMI** (*Post-mortem* Interval), * indicates the time after post-operative resection; **Testing Methods**: AFM: atomic force microscopy, IND: indentation, AMT: axial mechanical test, OST: oscillatory shear test, MRE: magnetic resonance elastography, USE: ultrasound elastography, PA: pipette aspiration ; **Loading Modes:** QSIT: quasi-static indentation test, UT: uniaxial tension, UC: uniaxial compression, SS: simple shear, TS: torsional shear, PS: pure shear, CT: creep test, SRT: stress relaxation test, CSRT: constant strain rate test, OLT: oscillatory loading test, CLT: cyclic loading test (large strain), FST: frequency sweep test, AST: amplitude sweep test, AT: adhesion test, MALT: multi-axial loading test, SWE: shear wave elastography, SE: strain elastography; **Specimen**: W: width, L: length, H: height, R:radius; **Temp** (temperature); **Fitted Material Models**: Hyperelastic: NH: neo-Hookean model, D: Demiray model, G: Gent model, MR: Mooney-Rivlin model, O: Ogden model ; Viscoelastic: Zn: Zener model (standard linear solid model), Mw: Maxwell model, Sp: springpot model, KV: Kelvin-Voigt model, Bg: Burgers model (four-parameter fluid model), PS: Prony series, LEV: linear viscoelastic model (linear), QLV: quasi-linear viscoelastic model (non-linear), GRV: Green-Rivlin viscoelastic model (non-linear), MD: multiplicative decomposition ; Poroelastic: TC: Terzaghi's Consolidation, BTP: Boit's theory of poroelasticity, DL: Darcy's law. **PIM** (Parameter Identification Method): LsF: Least-square Fit, IFEM: inverse identification using finite element modeling.

| Literatures | Regions | PMI (hours) | Ages (years) | Testing Methods | Loading Modes | Specimen (cm) | Temp (ºC) | Frequency/strain/ strain rate range | Measured Material Properties | Fitted Material Models | | | PIM |
|---|---|---|---|---|---|---|---|---|---|---|---|---|---|
| | | | | | | | | | | Hyperelastic | Viscoelastic | Poroelastic | |
| Fallenstein, et al. [206] | Cb | 10-62 | 44-92 | OST | OLT, SS | Rectangular W2 × L3 × H0.4-0.7 | 37 | 9-10 [Hz], 7-24.5%[-] | Storage modulus, Loss modulus | | KV | | LsF |
| Galford, et al. [208] | Cb | 6-12 | -- | AMT | CT, SRT, OLT | Cylindrical R0.318 × H0.635 | 37 | 10-40 [Hz] | Storage modulus, Loss modulus, Creep compliance, Relaxation modulus, Viscoelasticity | | Bg | | LsF |
| Estes, et al. [215] | CR | 7-12 | 52-84 | AMT | UC, CSRT | Cylindrical: R0.635 × H0.635 | 37 | 0-170% [-] 0.08-40 [$s^{-1}$] | Stress-strain curves | | | | LsF |
| Shuck, et al. [128] | CR, T | -- | -- | OST | FST,PS, OLT | Cylindrical: R0.635 × H1.27 | 37 | 5-350 [Hz], 1.3-3.5% [-] | Storage modulus, Loss modulus, Viscoelasticity, Limited strains & strain rates, Heterogeneity, Anisotropy | | Bg | | LsF |
| McElhaney, et al. [209] | Cb, | 6-10 | -- | IND | CSRT, QSIT | Cylindrical: R0.635 × H2.54 | 37 | 9-10 [Hz] | Bulk modulus, Viscosity | | | | LsF |
| Donnelly, et al. [210] | CC, Mb | < 48 | 44-92 | AMT | CSRT, SS | Cylindrical: R0.615-0.953 × H0.53-2.64 | 22 | 0-45% [-] 30-180 [$s^{-1}$] | Storage modulus, Viscosity, Viscoelasticity, Stress-strain curves, Heterogeneity, | | Zn | | LsF |
| Prange, et al. [216] | C | < 3* | -- | AMT | SS, SRT, UC | Rectangular: W0.5 × L1 × H0.1 | -- | 2.5-50% [-] 0.42-8.3 [$s^{-1}$] | Shear modulus, Anisotropy, Hyperelasticity, Heterogeneity, Viscoelasticity | O | PS | | LsF |
| Prange, et al. [41] | C | < 3* | -- | AMT | SS, SRT, UC | Rectangular: W0.5 × L1× H0.1 | -- | 2.5-50% [-] 0.42-8.3 [$s^{-1}$] | Hyperelasticity, Shear modulus, Anisotropy, Heterogeneity, Viscoelasticity | O | PS | | LsF |
| Takhounts, et al. [217] | Cb | < 24 | -- | AMT | SS, SRT, UC | Cylindrical: R2 × H0.9-1.8 | 22 | 12.5-50% [-] 3.125, 6.25 [$s^{-1}$] | Viscoelasticity | | LVE, QLV, GRV | | LsF |

(*Continued*)

**Table. 2. Literature summary of human brain tissue testing**. **Regions**: Cb: cerebrum (white and gray), C: cortex, T: thalamus, H: Hippocampus, CR: corona radiata, CC: corpus callosum, BG: basal ganglia, Mb: midbrain, BS: brain stem, CB: cerebellum, WB: whole brain, BT: brain tumor; **PMI** (*Post-mortem* Interval), * indicates the time after post-operative resection; **Testing Methods**: AFM: atomic force microscopy, IND: indentation, AMT: axial mechanical test, OST: oscillatory shear test, MRE: magnetic resonance elastography, USE: ultrasound elastography, PA: pipette aspiration ; **Loading Modes:** QSIT: quasi-static indentation test, UT: uniaxial tension, UC: uniaxial compression, SS: simple shear, TS: torsional shear, PS: pure shear, CT: creep test, SRT: stress relaxation test, CSRT: constant strain rate test, OLT: oscillatory loading test, CLT: cyclic loading test (large strain), FST: frequency sweep test, AST: amplitude sweep test, AT: adhesion test, MALT: multi-axial loading test, SWE: shear wave elastography, SE: strain elastography; **Specimen**: W: width, L: length, H: height, R:radius; **Temp** (temperature); **Fitted Material Models**: Hyperelastic: NH: neo-Hookean model, D: Demiray model, G: Gent model, MR: Mooney-Rivlin model, O: Ogden model ; Viscoelastic: Zn: Zener model (standard linear solid model), Mw: Maxwell model, Sp: springpot model, KV: Kelvin-Voigt model, Bg: Burgers model (four-parameter fluid model), PS: Prony series, LEV: linear viscoelastic model (linear), QLV: quasi-linear viscoelastic model (non-linear), GRV: Green-Rivlin viscoelastic model (non-linear), MD: multiplicative decomposition ; Poroelastic: TC: Terzaghi's Consolidation, BTP: Boit's theory of poroelasticity, DL: Darcy's law. **PIM** (Parameter Identification Method): LsF: Least-square Fit, IFEM: inverse identification using finite element modeling.

| **Literatures** | **Regions** | **PMI** (hours) | **Ages** (years) | **Testing Methods** | **Loading Modes** | **Specimen** (cm) | **Temp** (°C) | **Frequency/strain/ strain rate range** | **Measured Material Properties** | **Fitted Material Models** | | | **PIM** |
|---|---|---|---|---|---|---|---|---|---|---|---|---|---|
| | | | | | | | | | | Hyperelastic | Viscoelastic | Poroelastic | |
| Nicolle, et al. [112] | CR, T | > 72 | -- | OST | SRT,TS, OLT | Cylindrical: R0.5× H0.015-0.085; R1 × H0.225 | 37 | 0.1-10000 [Hz], 0.001% [-] | Hyperelasticity, Storage modulus, Loss modulus, Viscoelasticity, Anisotropy, Heterogeneity | O | Mw | | LsF |
| Franceschini, et al. [24] | C,CC,T | < 12 | -- | AMT | CSRT, UC, UT, CLT | Cylindrical: R0.7-0.75 × H0.8-0.98; Rectangular: W0.8 × L1.3 × H0.8 | 22 | 0-270% [-] 5.5-9.3e$^{-3}$ [s$^{-1}$] | Hyperelasticity, Hysteresis, Fracture & Damage, Viscoelasticity, Poroelasticity, Stress-strain curves | O | KV | TC | LsF |
| Schiavone, et al. [47] | C | -- | -- | PA | QSIT | -- | 37 | -- | Hyperelasticity | MR | | | IFEM |
| Zhu, et al. [25] | CC,T | >168 | 45 | AMT | UC | Rectangular: W1.5 × L1.5 × H0.8 | 37 | 0.5,5,35[s$^{-1}$] | Young's modulus, Shear modulus, Viscoelasticity, Heterogeneity, Stress-strain curves | | Zn | | IFEM |
| Chatelin, et al. [132] | CR, T, BS | 24-48 | 0.2 – 55 | OST | FST,TS, OLT | Cylindrical: R1cm × H0.2-0.5 cm | 37 | 0.1-10 Hz, 0.5% [-] | Storage modulus, Loss modulus, Heterogeneity | | LVE | | LsF |
| Jin, et al. [98] | C,T, CC,CR | ~96 | 45-94 | AMT | UT,UC,SS | Rectangular: W1.4 × L1.4 × H0.5 | 37 | 0-50% [-], 0.5-30 [s$^{-1}$] | Heterogeneity, Anisotropy, Stress-strain curves | | | | LsF |
| Forte, et al. [42] | Cb | 26-48 | 65-88 | OST, AMT | FST, OLT, SRT, UC, TS | Cylindrical: R1.25 × H0.2-0.8; R0.6 × H0.8 | 24; 37 | 0.01-25 Hz, 1% [-]; 0-35.6%, 1e$^{-4}$-1[s$^{-1}$] | Storage modulus, Loss modulus, Hyperelasticity, Viscoelasticity, Poroelasticity | O | PS | DL | LsF |
| Budday, et al. [30] | CC,CR,BG,C | < 60 | 54-81 | AMT | UT,UC,S,CLT,SRT,MALT | Rectangular: W0.5 cm × L0.5 cm × H0.5 cm | 22 | 0-10% [-]; 0-10% [-]; 0-20% [-]; | Hyperelasticity, Hysteresis, Anisotropy, Heterogeneity, Stress-strain curves | O,NH, MR,D, G | | | LsF |
| Budday, et al. [124] | CC,CR,BG,C | < 60 | 54-81 | AMT | UT,UC,SS,CLT,SRT,MALT | Rectangular: W0.5 × L0.5 × H0.5 | 22 | 0-10% [-]; 0-10% [-]; 0-20% [-]; | Viscoelasticity, Hyperelasticity, Hysteresis, Heterogeneity, Stress-strain curves | O | Zn | | LsF |

(*Continued*)

**Table. 2. Literature summary of human brain tissue testing**. **Regions**: Cb: cerebrum (white and gray), C: cortex, T: thalamus, H: Hippocampus, CR: corona radiata, CC: corpus callosum, BG: basal ganglia, Mb: midbrain, BS: brain stem, CB: cerebellum, WB: whole brain, BT: brain tumor; **PMI** (*Post-mortem* Interval), * indicates the time after post-operative resection; **Testing Methods**: AFM: atomic force microscopy, IND: indentation, AMT: axial mechanical test, OST: oscillatory shear test, MRE: magnetic resonance elastography, USE: ultrasound elastography, PA: pipette aspiration ; **Loading Modes:** QSIT: quasi-static indentation test, UT: uniaxial tension, UC: uniaxial compression, SS: simple shear, TS: torsional shear, PS: pure shear, CT: creep test, SRT: stress relaxation test, CSRT: constant strain rate test, OLT: oscillatory loading test, CLT: cyclic loading test (large strain), FST: frequency sweep test, AST: amplitude sweep test, AT: adhesion test, MALT: multi-axial loading test, SWE: shear wave elastography, SE: strain elastography; **Specimen**: W: width, L: length, H: height, R:radius; **Temp** (temperature); **Fitted Material Models**: Hyperelastic: NH: neo-Hookean model, D: Demiray model, G: Gent model, MR: Mooney-Rivlin model, O: Ogden model ; Viscoelastic: Zn: Zener model (standard linear solid model), Mw: Maxwell model, Sp: springpot model, KV: Kelvin-Voigt model, Bg: Burgers model (four-parameter fluid model), PS: Prony series, LEV: linear viscoelastic model (linear), QLV: quasi-linear viscoelastic model (non-linear), GRV: Green-Rivlin viscoelastic model (non-linear), MD: multiplicative decomposition ; Poroelastic: TC: Terzaghi's Consolidation, BTP: Boit's theory of poroelasticity, DL: Darcy's law. **PIM** (Parameter Identification Method): LsF: Least-square Fit, IFEM: inverse identification using finite element modeling.

| **Literatures** | **Regions** | **PMI** (hours) | **Ages** (years) | **Testing Methods** | **Loading Modes** | **Specimen** (cm) | **Temp** (ºC) | **Frequency/strain/ strain rate range** | **Measured Material Properties** | **Fitted Material Models** | | | **PIM** |
|---|---|---|---|---|---|---|---|---|---|---|---|---|---|
| | | | | | | | | | | Hyperelastic | Viscoelastic | Poroelastic | |
| Budday, et al. [125] | CC,CR,BG,C | < 60 | 54-81 | AMT | UT,UC,SS,CLT,SRT,MALT | Rectangular: W0.5 × L0.5 × H0.5 | 22 | 0-10% [-]; 0-10% [-]; 0-20% [-]; | Viscoelasticity, Hyperelasticity, Hysteresis, Heterogeneity, Stress-strain curves | O | PS, Zn | | LsF |
| Finan, et al. [211] | C, H | < 6* | 4-58 | IND | SRT | H0.1 | 22 | 0-10% [-],$5e^{-3}$-5 [$s^{-1}$] | Shear modulus, Viscosity, Viscoelasticity, Heterogeneity | | PS | | LsF |
| Stewart, et al. [76] | BT | 3-4* | -- | IND | SRT | H0.3 | 37 | 0-10% [-] | Shear modulus, Viscoelasticity | | Zn | | LsF |
| Karimi, et al. [213] | CB | <10 | 60-80 | AMT | UC, CLT | Proper size | 37 | 0-50% [-], 1[$s^{-1}$] | Young's modulus, Failure stress, Stress strain curve | | | | LsF |
| Park, et al. [212] | C | 2.5-14.5 | 61-75 | AFM | FS, QSIT | H0.0008 | 21 | 0.01-10 Hz, 0-2% [-], | Young's modulus, Storage modulus, Loss modulus, Hysteresis, Surface roughness, Viscosity, Stress-strain curve | | KV | | LsF |
| Budday, et al. [114] | CC,CR,BG,C | 24-60 | 55-68 | AMT | SS, SRT, UC, UT, CLT | Rectangular: W0.5 × L0.5 × H0.5 | 22 | 0-20% [-], 0.0067 [$s^{-1}$]; 20% [-], 0.33 [$s^{-1}$]; 0-10% [-]; 0-10% [-] | Viscoelasticity, Viscosity, Hyperelasticity, Hysteresis, Heterogeneity, Stress-strain curves | O | PS | | LsF |
| MacManus, et al. [33] | C, CB, BS | <96 | 64-94 | IND | SRT | Cylindrical: R0.125 × H0.125 | 22 | 0-30% [-], 10 [$s^{-1}$] | Shear modulus, Viscoelasticity, Hyperelasticity, Heterogeneity | NH | PS | | IFEM |
| Menichetti, et al. [28] | CC,CR,BG,C,BS, | <96 | 64-94 | IND | SRT | H2 | 37 | 0-35% [-]; 10 [$s^{-1}$] | Shear modulus, Relaxation modulus, Viscoelasticity, Hyperelasticity, Heterogeneity, Heterogeneity | NH | PS | | IFEM |
| Greiner, et al. [93] | C, CR | -- | 77 | IND, AMT | QSIT, UT, UC, SRT, CLT | Cylindrical: R0.4 × H0.4 | 37 | 0-15% [-] | Hyperelasticity, Viscoelasticity, Poroelasticity, Stress-strain curve | O | MD | DL | |

(*Continued*)

**Table. 2. Literature summary of human brain tissue testing**. **Regions**: Cb: cerebrum (white and gray), C: cortex, T: thalamus, H: Hippocampus, CR: corona radiata, CC: corpus callosum, BG: basal ganglia, Mb: midbrain, BS: brain stem, CB: cerebellum, WB: whole brain, BT: brain tumor; **PMI** (*Post-mortem* Interval), * indicates the time after post-operative resection; **Testing Methods**: AFM: atomic force microscopy, IND: indentation, AMT: axial mechanical test, OST: oscillatory shear test, MRE: magnetic resonance elastography, USE: ultrasound elastography, PA: pipette aspiration ; **Loading Modes:** QSIT: quasi-static indentation test, UT: uniaxial tension, UC: uniaxial compression, SS: simple shear, TS: torsional shear, PS: pure shear, CT: creep test, SRT: stress relaxation test, CSRT: constant strain rate test, OLT: oscillatory loading test, CLT: cyclic loading test (large strain), FST: frequency sweep test, AST: amplitude sweep test, AT: adhesion test, MALT: multi-axial loading test, SWE: shear wave elastography, SE: strain elastography; **Specimen**: W: width, L: length, H: height, R:radius; **Temp** (temperature); **Fitted Material Models**: Hyperelastic: NH: neo-Hookean model, D: Demiray model, G: Gent model, MR: Mooney-Rivlin model, O: Ogden model ; Viscoelastic: Zn: Zener model (standard linear solid model), Mw: Maxwell model, Sp: springpot model, KV: Kelvin-Voigt model, Bg: Burgers model (four-parameter fluid model), PS: Prony series, LEV: linear viscoelastic model (linear), QLV: quasi-linear viscoelastic model (non-linear), GRV: Green-Rivlin viscoelastic model (non-linear), MD: multiplicative decomposition ; Poroelastic: TC: Terzaghi's Consolidation, BTP: Boit's theory of poroelasticity, DL: Darcy's law. **PIM** (Parameter Identification Method): LsF: Least-square Fit, IFEM: inverse identification using finite element modeling.

| **Literatures** | **Regions** | **PMI** (hours) | **Ages** (years) | **Testing Methods** | **Loading Modes** | **Specimen** (cm) | **Temp** (ºC) | **Frequency/strain/ strain rate range** | **Measured Material Properties** | **Fitted Material Models** | | | **PIM** |
|---|---|---|---|---|---|---|---|---|---|---|---|---|---|
| | | | | | | | | | | Hyperelastic | Viscoelastic | Poroelastic | |
| Pan, et al. [19] | C | <3* | 35-52 | IND | SRT | H0.6 | 22 | 0-17% [-] | Shear modulus, Viscoelasticity | | QLV | | LsF |
| Sundaresh, et al. [218] | C, H | < 6* | 4-58 | IND | SRT | H0.1 | -- | 0-30% [-], 1.9 [$s^{-1}$] | Shear modulus, Hyperelasticity, Viscoelasticity, Heterogeneity | NH, MR, O | QLV | | LsF |
| Hinrichsen, et al. [26] | CC, CR, C, BGH, Mb, BS, T, CB | <72 | 62-92 | AMT | UT,UC,TS,SRT,FS,CLT | Cylindrical: R0.4 × H0.27-0.72 | 37 | 0-15% [-], 0-30% [-], 1 $e^{-2}$ [$s^{-1}$] | Shear modulus, Bulk modulus Hyperelasticity, Hysteresis, Heterogeneity, Stress-strain curves | O | | | IFEM |
| Su, et al. [214] | BT | <12 | 72 | AMT | UC,SRT | Cylindrical: R0.5 × H0.46 | 22 | 0-10% [-],1 [$s^{-1}$] | Young's modulus, Relaxation modulus, Poisson's ratio, Hydraulic permeability, Viscoelasticity, Poroelasticity | | PS | BTP | LsF |
| Basilio, et al. [94] | C,H | <6* | 4-58 | IND | SRT | H0.1 | 22 | 0-30% [-], 0.79-3.57 [$s^{-1}$] | Hyperelasticity, Viscoelasticity, Heterogeneity | NH, MR, O | QLV | | IFEM |
| Skambath, et al. [77] | BT,C | < 0.1 | -- | IND | QSIT | H0.2-1 | 20 | 0-10% [-] | Young's modulus | | | | LsF |
| Greiner, et al. [207] | C,CR | | 57-71 | AMT | UT,UC,SRT,CLT | Cylindrical: R0.4 × H0.27-0.72 | 37 | 0-15% [-], 1$e^{-2}$[$s^{-1}$] | Viscoelasticity, Heterogeneity, Poroelasticity | | MD | DL | IFEM |

In addition to the global comparison between gray and white matter, we also summarized the specific shear stiffness values reported for individual brain regions, as illustrated in Figure 5. Here, only commonly dissected regions were included, and for studies that measured multiple anatomical subregions, values were averaged again across regions corresponding to the broader parcellation stated here. For example, Menichetti, et al. [28] reported viscoelastic properties for 12 distinct anatomical regions of the human brain, including six subregions within the cortex and two within the brain stem. Another study by Hinrichsen, et al. [26] measured the mechanical properties of 19 anatomical human brain regions, including five cortex subregions, three basal ganglia subregions, and two subregions each from brain stem, corona radiata, cerebellum, and midbrain, respectively. As shown in Figure 5, considerable variability exists even within these localized anatomical regions, particularly for the cerebellum, corona radiata, corpus callosum, basal ganglia, and cortex. For other regions, smaller deviations are observed, however, this is likely attributed to the limited number of studies rather than consistency in the mechanical properties. Further investigation is needed to determine whether these smaller deviations are representative or simply the result of insufficient data.

The substantial variability observed in reported brain tissue mechanical properties arises from a combination of factors. As discussed in Section 2, these include differences in subject-specific variables such as age and gender, tissue sample preparation (e.g., post-mortem time, sampling size, hydration level, anatomical location, and sampling direction), experimental conditions (e.g., temperature, preconditioning, and loading rate or frequency for dynamic tests), and the choice of constitutive model (e.g., linear, hyperelastic, viscoelastic, or poroelastic), all of which may affect the testing outcomes and data interpretation. To visualize the scope of these influences, we compiled the potential effects reported in each study into Figure 6. In this figure, each factor is marked as “1” if considered influential and “0” if deemed negligible. While these effects have been comprehensively reviewed in prior literature [21, 45, 71, 219], here we provide a brief discussion based on the summary of human brain tissue testing studies.

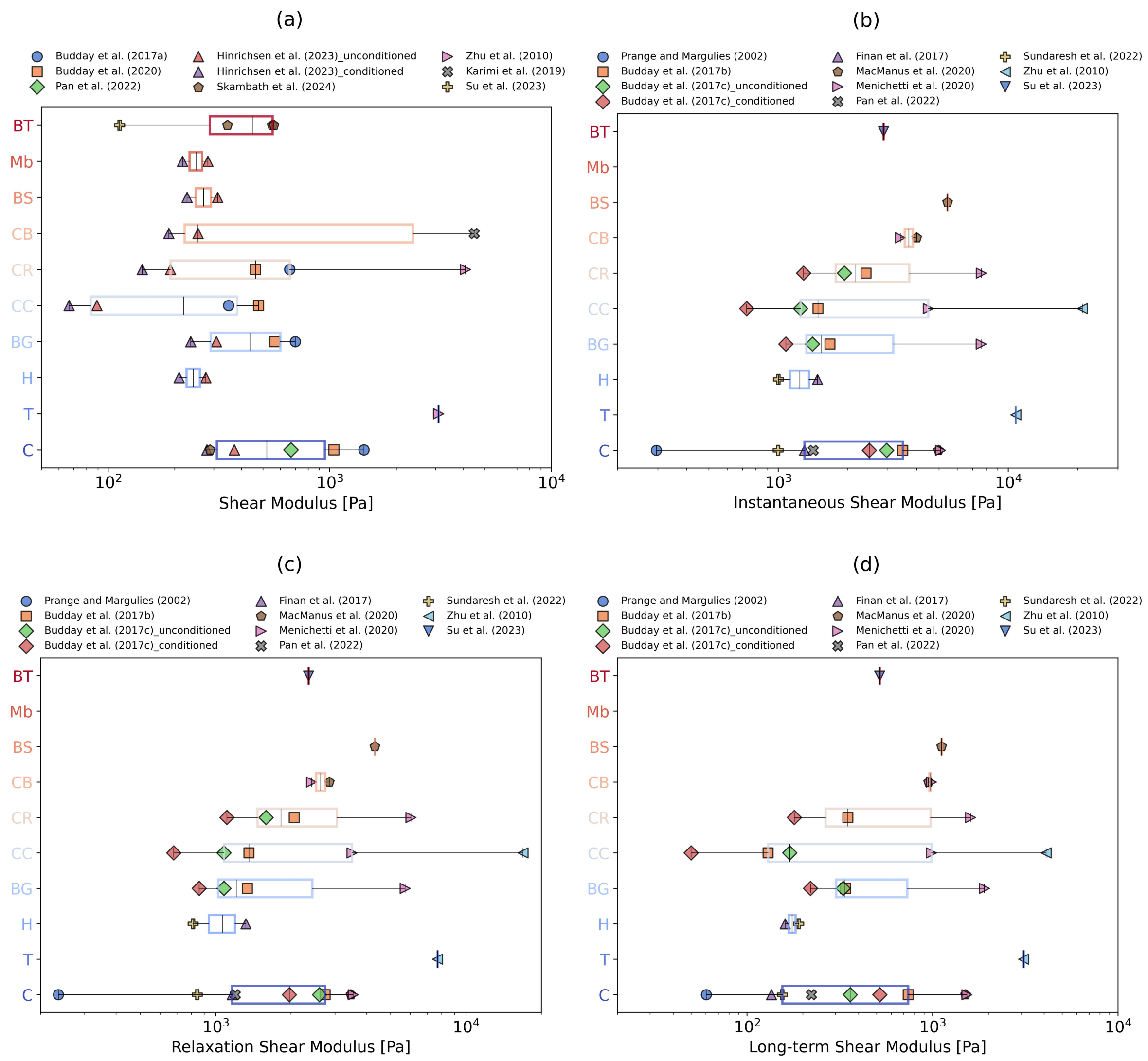

**Figure 5. Summary of various shear moduli for human brain regions**. Box plot and data points illustrate the distribution of regional shear modulus (a), instantaneous shear modulus (b), relaxation shear modulus (c), and long-term shear modulus (d) based on collected in-vitro or ex-vivo studies, as presented in Tabel 2. C: cortex, T: thalamus, H: Hippocampus, BG: basal ganglia, CC: corpus callosum, CR: corona radiate, CB: cerebellum, BS: brain stem, Mb: midbrain, BT: brain tumor.

*Age and gender effects on human brain mechanics*

Throughout development, maturation, and aging, the brain undergoes significant structural and compositional changes, implying that age may influence its mechanical properties. This effect has been well-documented in animal studies, where mechanical testing has shown clear age-dependent trends [34, 41, 134]. However, in human brain tissue, aging appears to have a minor influence on mechanical properties [28, 30, 94, 98, 211, 218]. Most of these studies focused on adult brain tissue, with small age variations across limited samples. A notable exception is the study by Finan, et al. [211], who conducted an IND test on 11 human brains ranging from 4 to 58 years of age, covering children, adolescents, and adults. Despite this age range, no significant correlation was found between age and the shear modulus of cortical gray or white matter. Only one study by Chatelin, et al. [132] reported a notable age-related effect. They performed OST on both human child brains (five subjects aged 5 to 22 months) and adult brains (two subjects aged 50 to 55 years). Their results identified a significant increase in both storage and loss moduli with age in the pediatric group and found adult brain tissue to be three to four times stiffer than the younger brains. It is worth noting that the conclusion made on the aging effect may lack statistical robustness, as the small sample size used in their studies due to the inherent challenges of acquiring human brain tissue across a broad age range. In contrast, the noninvasive approach like MRE offers a more practical and ethical approach for assessing the aging effect *in vivo*, and several studies have consistently reported age-related stiffening in the human brain using MRE [138, 140, 161]. Similar challenges exist in investigating the influence of gender on brain mechanics. Amony limited evidence, Finan, et al. [211] reported a significant gender difference in the stress relaxation behaviors of cortical white matter. Specifically, male brain tissue exhibited greater modulus decay during stress relaxation, although long-term shear modulus values seem consistent between genders.

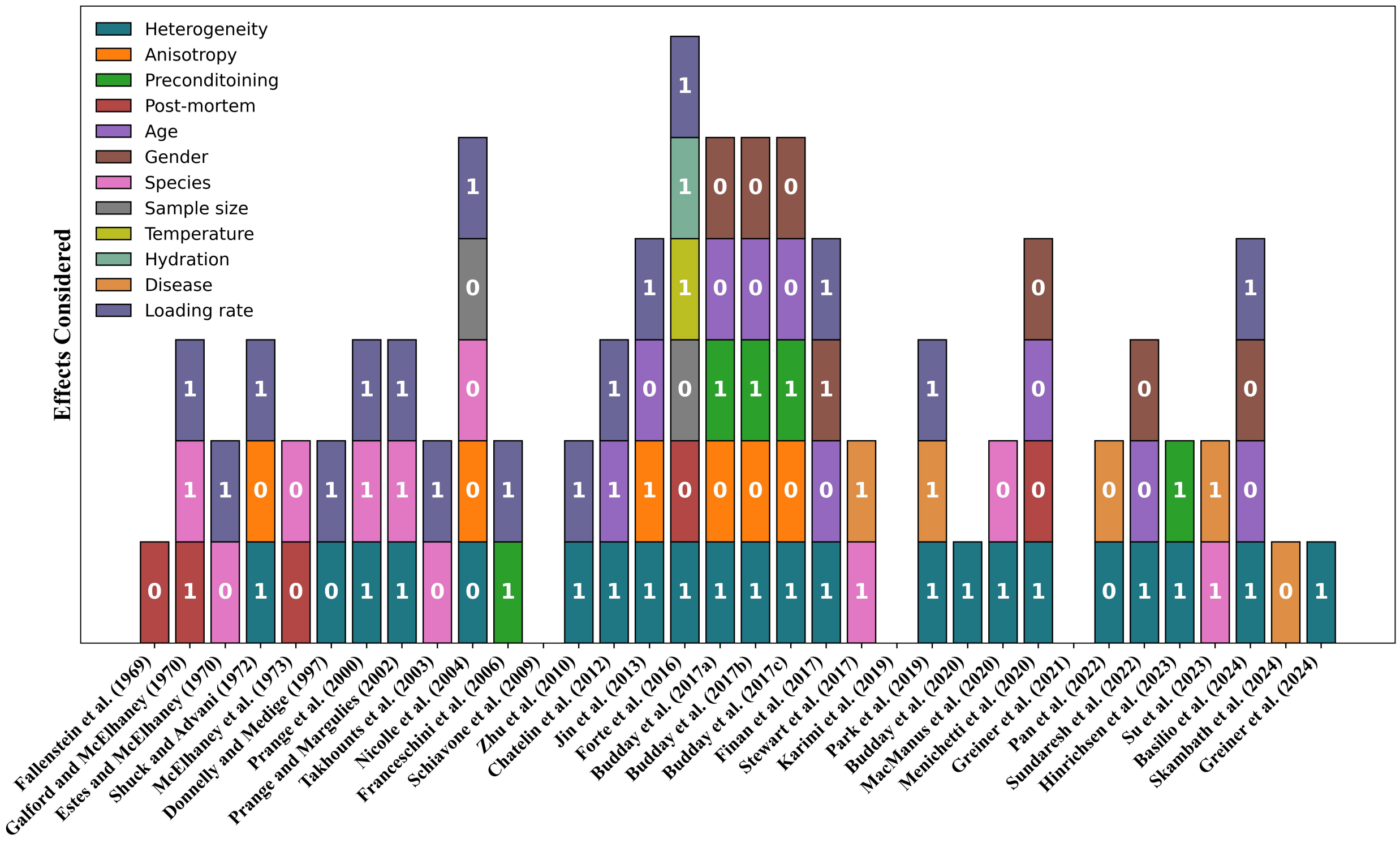

**Figure 6. Overview of effects considered in human brain tissue mechanical testing across various studies**. The colored blocks represent the effects investigated and discussed in each publication. The in-block text "1" indicates that the effect may significantly influence testing results, while "0" suggests minimal impact.

*Predominance of mechanical heterogeneity over anisotropy in the human brain*

Given the intricate anatomical structure of the human brain, sample preparation has a non-negligible impact on tissue mechanical testing outcomes. Numerous studies have consistently reported mechanical heterogeneity across different human brain regions. However, the human brain generally exhibits only minor anisotropy (i.e., direction-dependent behavior) in its mechanical properties—even within the corpus callosum, which contains densely aligned axonal fiber tract [30, 112, 128]. For instance, Budday, et al. [30] investigated the directional mechanical response of the corona radiata and corpus callosum. In their tests, compression and tension tests were conducted both along and perpendicular to fiber orientation, while simple shear tests in three distinct directions relative to the fiber alignments. Their results showed minimal directional dependence, although the tissue was slightly softer in compression and marginally stiffer in tension when tested along the fiber. In contrast, Jin, et al. [98] reported significantly higher shear stress when loading was applied along the fiber direction compared to transverse loading in white matter. However, no anisotropy was observed in their tension and compression results. Notably, the apparent anisotropy might have stemmed from dimensional effects, as Jin's study used rectangular specimens with unequal lateral dimensions (14 × 14 × 5 mm), which could have influenced the outcomes Budday, et al. [30].

*Difference caused by sample size and post-mortem interval*

To facilitate mechanical testing, brain tissue samples are often prepared in either rectangular or cylindrical shapes, with their dimensions ranging from a few millimeters to several centimeters, especially in studies utilizing OST and AMT (see Table 2). However, sample size has not been found to significantly influence mechanical outcomes in the two existing human brain studies on this topic [42, 112]. Both studies used cylindrical specimens of varying thickness for OST and reported that thickness has little to no effect on the result, once the samples were securely affixed to the testing plates. While this suggests that OST measurements are size-independent within small deformation ranges, further investigation is warranted to determine whether varying sample dimensions introduce bias in larger deformation tests such as AMT, which potentially leads to artificial anisotropy artifacts. Another concern related to the size effect is the inconsistency between the sample's intended size during preparation and its actual size during testing, particularly for large samples. Due to the brain's ultrasoft nature, significant deformation can occur under gravitational loading alone [21]. One approach to address this issue is inverse parameter identification via FEM. A more precise strategy was proposed by Zhu, et al. [25], who employed a laser scanning system in conjunction with a 3D surface reconstruction algorithm to capture the realistic geometry of each brain sample prior to testing and FEM simulation. Post-mortem interval is another critical factor to consider in invasive mechanical tests. Although brain tissue experiences structural variations due to biochemical degradation, enzymatic activity, and water content rapidly after death, several studies have reported that the effect of post-mortem appears to be negligible when tissue is appropriately preserved [21, 112, 206, 209]. For example, Menichetti, et al. [28] found no significant impact of post-mortem delay on inter-regional mechanical difference, and Forte, et al. [42] similarly concluded that varying post-mortem durations between 26 and 48 hours did not affect mechanical outcomes in their study population.

*Undervalued effects of temperature and humidity*

Humidity and temperature controls are also important for human brain tissue mechanical testing, especially during long-duration experiments such as OST or quasistatic AMT. Forte, et al. [42] systematically examined the influence of both factors on OST results. They evaluated humidity effects by testing brain tissue under three conditions: continuous water misting for full humidity control, no moisture regulation, and periodic rehydration using saline. These tests were conducted at controlled temperatures of 24 ºC and 37 ºC. For temperature effects, a sweep from 22 ºC to 37 ºC was performed, both with and without humidity control. Their findings revealed that in the absence of humidity control, tissue dehydrates rapidly and stiffens significantly. This phenomenon is more notable at higher temperatures (37 ºC), where the measured storage modulus increased by up to 21.9 times. Partial recovery of stiffness was observed after rehydration. In temperature sweep analysis, a strong stiffening trend emerged when no moisture control was applied, whereas the opposite trend was observed under full humidity control, with both storage and loss moduli decreasing by 1.4 and 1.6 times at 37 ºC compared to 22 ºC, respectively. Despite these findings, many studies on human brain tissue mechanics have been conducted at room temperature (22 or 24 ºC, see Table 2) rather than at physiological temperature (37 ºC), often overlooking the effects of temperature [30, 33, 94, 211, 214]. This oversight is likely based on early evidence derived from animal brain tests [106], though these were based only on discrete comparisons at 22 ºC and 37 ºC, not a continuous temperature sweep like in the study of Forte, et al. [42]. Therefore, to reduce potential artifacts introduced by temperature variability, it is recommended to maintain environmental temperature close to physiological conditions (37 °C) during mechanical testing of human brain tissue.

*Loading rate–dependent behavior of human brain tissue*

In dynamic testing of the human brain tissue, viscoelasticity-related measures such as storage and loss modulus, are commonly found to be frequency dependent, as shown in Figures 7(a)-(b). Regardless of differences in testing methods, loading modes, or deformation levels, a consistent trend was observed that both the elastic (storage) and viscous (loss) resistance of the brain increased with frequency, reflecting enhanced stiffness and energy dissipation at higher loading rates. Analogously, as demonstrated in Figures 7(c)-(d), the human brain exhibits pronounced compression or shear softening under prolonged loading, underscoring its time-dependent mechanical behavior. These viscoelastic attributes highlight the strong sensitivity of the brain mechanical response to the rate of applied loading. Figures 8 further illustrate this rate dependence by collecting stress-strain data reported from various AMT studies on the human brain, with Figures 8(a)-(c) focusing on small deformation cases ($\epsilon_{\max} < 0.25$) and Figures 8(d)-(f) on large deformation range ($\epsilon_{\max} > 0.25$). The tested brain regions and corresponding loading rates are also present in the figure legend for inter- and intra-study comparison. As seen in the Figure 8, the human brain exhibits a pronounced strain rate stiffening behavior. For example, in the study by Zhu, et al. [25], the maximum stress in the corona radiata at 65 % strain was nearly 2.5 times higher when the strain rate increased from 0.8 $s^{-1}$ to 40 $s^{-1}$, emphasizing the importance of loading rate control in accurately characterizing brain mechanics.

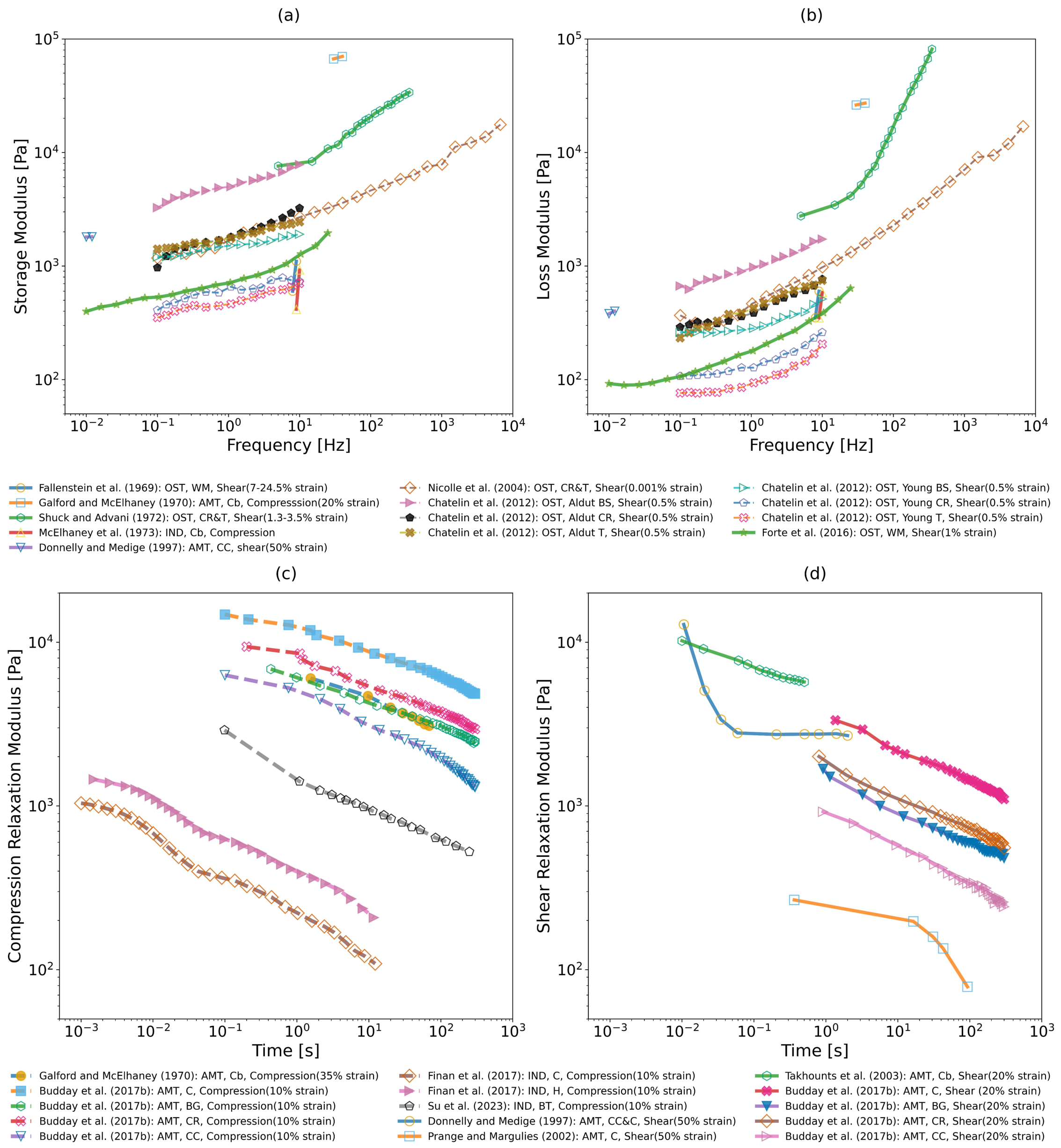

**Figure 7. Summary of various moduli reported in the literature for human brain tissue**. Frequency-dependent storage modulus (a) and loss modulus (b); time-dependent compressive relaxation modulus (c) and shear relaxation modulus (d). The data references are provided along with the details on the testing methods, tested regions, loading modes, and the applied strains.

*Conditioning effects on human brain mechanics*

The preconditioning effect is also evident in human brain tissue mechanical testing. After conditioning, brain tissue responses tend to become more stable and repeatable. Therefore, the data recorded during post-conditioning cycles—typically the second or third—are frequently used for calibrating constitutive models [24, 30, 41]. The conditioning effects is generally attributed to microstructural adjustments or minor damage occurring during the initial loading cycles.

Interestingly, preconditioning behavior appears to be recoverable under small deformations. For instance, Budday, et al. [30] observed that brain tissue, after resting for one hour, could fully recover and display a similar preconditioning response as in the initial test. Based on this, they attributed this effect to reversible changes in tissue state—such as interstitial fluid redistribution or recoverable intracellular interactions—rather than the irreversible microstructural damage like microstructural reorganization, owing to the porous and fluid-saturated nature of brain tissue. Moreover, Budday, et al. [125] emphasized that data collected during the preconditioning phase are also valuable as reflections of *in vivo* physiological conditions, while the conditioned data can serve as reproducible baseline for *ex vivo* mechanical testing. Similarly, all of the reported stress-strain data, as depicted in Figure 8, are valuable regarding different research and clinical objectives. Data obtained under quasistatic scenarios within the small deformation range are suitable for representing long-term brain behaviors, such as those associated with brain development, aging, or disease progression. In contrast, stress-strain data measured under large deformation and high loading rate are critical for capturing the brain's mechanical response over shorter timescales, as encountered in scenarios such as TBI.

*Modeling assumptions in human brain mechanics*

Human brain tissue mechanical characterizations are also influenced by the assumptions embedded in the constitutive models used to calibrate material parameters. Assumptions such as isotropy versus anisotropy, incompressibility versus compressibility, and the selection of material type—elastic, hyperelastic, viscoelastic, or poroelastic—can all lead to different interpretations of the same experimental data [26]. As summarized in Table 2, various constitutive models have been applied in current studies to capture the brain's complex mechanical responses. For long-time scale, such as modeling brain development, hyperelastic models—particularly the Ogden-type models—are well-suited for capturing nonlinear elastic behavior [26, 30]. Conversely, at shorter timescales where the time-dependent effects become dominant, viscoelastic or poro-viscoelastic constitutive models provide a more accurate representation [24, 42]. For example, Greiner, et al. [93] modeled brain tissue as a poro-viscoelastic medium, where the solid matrix—including the network of cells embedded within the extracellular matrix—accounts for the viscoelastic contribution, and the free-flowing interstitial fluid contributes to the poroelastic effect. Through parameter studies, they emphasized that the brain's nonlinear behaviors cannot be captured by a single effective modulus, as derived from simple indentation tests. Instead, a combination of cyclic and stress relaxation experiments across multiple loading modes is necessary for reliable calibration of viscoelastic parameters. They further attributed discrepancies between compression and indentation results to the intricate interplay of poroelastic and viscous effects with inherent material nonlinearities. In a follow-up study, Greiner, et al. [207] proposed a six-parameter poro-viscoelastic framework based on multiplicative decomposition, which successfully captured the brain's combined responses to cyclic tension, compression, and relaxation. Additionally, a pilot study by Su, et al. [214] proposed a novel scaling approach to separate poroelastic and viscoelastic contributions across time scale. By scaling the relaxation force and time with the square of the sample length, they revealed a clear transition point between the viscoelasticity-dominated short-time regime and the poroelasticity-dominated long-time regime.

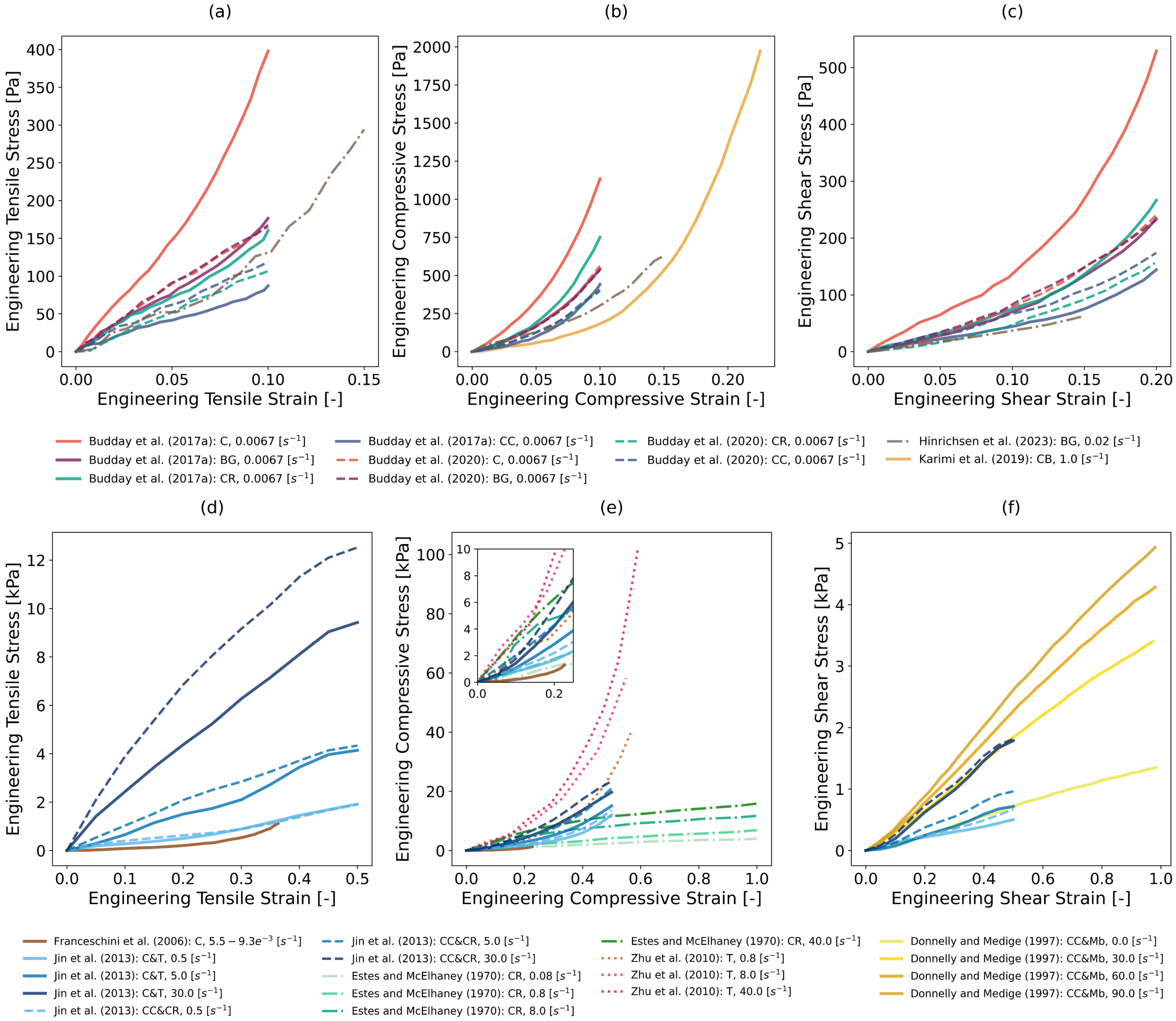

**Figure 8. Stress-strain curves of the human brain tissue under small and large deformation**. (a)-(c) represent the tension, compression, and shear data measured within small deformation range. (d)-(f) represent the tension, compression, and shear data measured within large deformation range. The inset in (e) provides a zoom-in view of recorded curves for the first 20% strain. All testing data were obtained through axial mechanical testing. The data references including details on the tested brain regions and applied strain rates are present at the bottom of the figure.

### 3.2. Noninvasive mechanical testing: in vivo, accessible, and diagnostic characterization

For the noninvasive studies, we primarily focused on two techniques: MRE and USE. Given the substantial body of literature in this field, we selectively reviewed representative studies that offer meaningful insight into the potential of noninvasive methods for estimating brain mechanical properties. These studies are summarized in Table 3, which includes key details such as the number of subjects, age and gender distribution, interested brain regions, testing methods, frequencies used, reported shear modulus values, and the factors considered like the age, gender, and pathological conditions. In the following section, we briefly discuss the findings from these noninvasive studies, with a primary focus on their comparison to invasive testing results presented in Section 3.1. This comparison serves as the basis for a broader discussion on the difference between *in vivo* and *ex vivo* assessments of brain tissue mechanical properties.

Compared to invasive testing, noninvasive testing generally includes more participants, often including dozens or hundreds of participants (see Table 3). This is primarily due to the key advantages of noninvasive, which can assess brain mechanical properties *in vivo* without the need for surgical intervention or tissue extraction. As a result, these methods pose minimal risk and discomfort to participants, making recruitment easier and more ethically feasible, particularly in healthy populations. The availability of large datasets enables more robust statistical analyses, thereby improving the reliability and generalizability of findings across various populations and conditions. Additionally, noninvasive testing is capable of characterizing regional brain properties. Using an accurate anatomical atlas, techniques such as MRE are able to quantify material properties in small or deep-located brain structures, which are significantly challenging in sample preparation for *ex vivo* testing. For example, McIlvain, et al. [140] utilized high-resolution MRE to investigate regional mechanical properties in both pediatric and adult brains across an age range of five to 35 years. Through anatomical parcellation, they successfully mapped age-related trajectories of stiffness and damping ratio in multiple brain regions, including finely parcellated cortical areas and deep structures such as the Hippocampus, Caudate, and Amygdala. Their findings highlighted distinct maturation patterns in mechanical properties across different regions.

**Table 3 Literature summary of noninvasive human brain tissue testing**. **N**: subject number, female and male number are also presented; **Regions**: WB: whole brain, G: gray matter, W: white matter, C: cortex, T: thalamus, H: Hippocampus, CC: corpus callosum, BS: brain stem, CB: cerebellum, BT: brain tumor; **Testing Methods**: MRE: magnetic resonance elastography, USE: ultrasound elastography; **Glo/Dis**: shear modulus of the global brain tissue or diseased brain tissue, *f25* indicates the modulus measured at the frequency of 25 Hz; **Age**: effect of age considered in the study,"0": no significant effect, "-": negative correlation (decrease with age), "+": positive correlation; **Gender**: gender effect considered in the study, (F > M): shear stiffness of female brain greater than male brain; "0": no significant gender difference; Disease: diseased brain tissue been tested, BT: brain tumor, AD: Alzheimer's disease, E: Epilepsy, PD: Parkinson's disease, D: Dementia, NPH: Normal pressure hydrocephalus.

| **Literature** | **N** (F/M) | **Age** (years) | **Regions** | **Testing Methods** | **Frequency** (Hz) | **Shear Modulus [kPa]** | | | **Effects** | | |
|---|---|---|---|---|---|---|---|---|---|---|---|
| | | | | | | Glo/Dis | Gray | White | Age | Gender | Disease |
| Kruse, et al. [151] | 25 | 23-79 | G,W | MRE | 100 | | 5.22 | 13.6 | √ (0) | | |
| Sack, et al. [161] | 55 (24/31) | 18-88 | WB | MRE | 25-62.5 | *f25*: 1.21<br>*f37.5*: 1.43<br>*f50*: 1.63<br>*f62.5*: 2.16 | | | √ (-) | √ (F > M) | |
| Sack, et al. [159] | 66 (35/31) | 18-72 | WB | MRE | 25-62.5 | *f25*: 1.82<br>*f37.5*: 2.18<br>*f50*: 2.39<br>*f62.5*: 2.89 | | | √ (-) | √ (F > M) | |
| Weaver, et al. [153] | 6 (2/4) | 25-55 | WB,W,G | MRE | 100 | 2.34 | 2.14 | 2.40 | | | |
| Guo, et al. [220] | 23 | 22-72 | W,CC,T | MRE | 30-60 | | 1.06 | 1.25 | | | |
| Johnson, et al. [163] | 7 (0/7) | 24-53 | CC,CR | MRE | 50 | | 2.27 | 3.07 | | | |
| Murphy, et al. [95] | 10 (2/8) | 23-55 | C,CB | MRE | 60 | 2.99 | 3.10 | | | | |
| Simon, et al. [169] | 16 (11/5) | 26-78 | BT | MRE | 45 | 1.40 | | 1.83 | | | √ (BT) |
| Braun, et al. [164] | 5 (0/5) | 26-55 | G,W | MRE | 40-60 | | 0.89 | 1.08 | | | |
| McGarry, et al. [156] | 2 (0/2) | 24,51 | G,W | MRE | 1,50 | | 2.20 | 2.80 | | | |
| Su, et al. [195] | 41 (19/22) | neonates | W,T,CB | USE | 3.5e6 | | | | √ (+) | | |
| Huston III, et al. [174] | 5 (0/5) | 53-65 | WB, C, CB | MRE | 60 | 2.77/2.59 | 2.91 | | | | √ (D) |

(*Continued*)

**Table 3 Literature summary of noninvasive human brain tissue testing**. **N**: subject number, female and male number are also presented; **Regions**: WB: whole brain, G: gray matter, W: white matter, C: cortex, T: thalamus, H: Hippocampus, CC: corpus callosum, BS: brain stem, CB: cerebellum, BT: brain tumor; **Testing Methods**: MRE: magnetic resonance elastography, USE: ultrasound elastography; **Glo/Dis**: shear modulus of the global brain tissue or diseased brain tissue, *f25* indicates the modulus measured at the frequency of 25 Hz; **Age**: effect of age considered in the study,"0": no significant effect, "-": negative correlation (decrease with age), "+": positive correlation; **Gender**: gender effect considered in the study, (F > M): shear stiffness of female brain greater than male brain; "0": no significant gender difference; Disease: diseased brain tissue been tested, BT: brain tumor, AD: Alzheimer's disease, E: Epilepsy, PD: Parkinson's disease, D: Dementia, NPH: Normal pressure hydrocephalus.

| **Literature** | **N** (F/M) | **Age** (years) | **Regions** | **Testing Methods** | **Frequency** (Hz) | **Shear Modulus [kPa]** Glo/Dis | Gray | White | **Effects** Age | Gender | Disease |
|---|---|---|---|---|---|---|---|---|---|---|---|
| Chauvet, et al. [196] | 63 | 24-85 | BT | USE | 9e6 | 11.01,7.9, 3.82,5.57 | | | | | √ (BT) |
| Murphy, et al. [172] | 48 (22/26) | -- | WB, C, CB | MRE | 60 | 2.51/2.40 | 2.65 | | | | √ (AD) |
| Kim, et al. [191] | 21 | neonates | G,W | USE | 3-16e6 | | | | √ (+) | | |
| Lipp, et al. [173] | 59 (24/35) | 49-82 | WB,C,T | MRE | 30-60 | 1.04/0.96 | 1.06 | | | | √ (PD) |
| Albayrak, et al. [192] | 83 (42/41) | neonates | G,W,T | USE | 1-6e6 | | 8.58 | 6.81 | √ (+) | √ (0) | |
| Tzschätzsch, et al. [96] | 26 (9/17) | 21-86 | WB | USE | 27-56 | 2.44 | | | | | |
| Huang, et al. [137] | 10 (4/6) | 24-38 | WB,G,W | MRE | 40-60 | *f40*: 2.57<br>*f50*: 3.04<br>*f60*: 3.27 | *f40*: 2.24<br>*f50*: 2.82<br>*f60*: 3.33 | *f40*: 3.36<br>*f50*: 3.78<br>*f60*: 3.85 | | | |
| Dirrichs, et al. [200] | 184 | neonates | | USE | 1.5e7 | | | | | | √ (NPH) |
| Yeung, et al. [138] | 36 | 7-44 | GW | MRE | 30-60 | | *f30*: 1.07<br>*f40*: 1.50<br>*f60*: 2.21 | *f30*: 1.12<br>*f40*: 1.54<br>*f60*: 2.24 | √ (0) | | |
| Huesmann, et al. [176] | 12 (10/2) | 26-61 | H | MRE | 50 | | | | | | √ (E) |
| Smith, et al. [177] | 4 (1/3) | 2-32 | CC | MRE | 50 | | | 3.78 | | | |
| Ozkaya, et al. [162] | 26 (13/13) | 7-17 | WB,G,W | MRE | 40-80 | *f40*: 1.69<br>*f60*: 2.37<br>*f80*: 2.75 | *f40*: 1.65<br>*f60*: 2.35<br>*f80*: 2.74 | *f40*: 1.83<br>*f60*: 2.45<br>*f80*: 2.76 | √ (0) | √ (F > M) | |

(*Continued*)

**Table 3 Literature summary of noninvasive human brain tissue testing**. **N**: subject number, female and male number are also presented; **Regions**: WB: whole brain, G: gray matter, W: white matter, C: cortex, T: thalamus, H: Hippocampus, CC: corpus callosum, BS: brain stem, CB: cerebellum, BT: brain tumor; **Testing Methods**: MRE: magnetic resonance elastography, USE: ultrasound elastography; **Glo/Dis**: shear modulus of the global brain tissue or diseased brain tissue, *f25* indicates the modulus measured at the frequency of 25 Hz; **Age**: effect of age considered in the study,"0": no significant effect, "-": negative correlation (decrease with age), "+": positive correlation; **Gender**: gender effect considered in the study, (F > M): shear stiffness of female brain greater than male brain; "0": no significant gender difference; Disease: diseased brain tissue been tested, BT: brain tumor, AD: Alzheimer's disease, E: Epilepsy, PD: Parkinson's disease, D: Dementia, NPH: Normal pressure hydrocephalus.

| **Literature** | **N** | **Age** | **Regions** | **Testing** | **Frequency** | **Shear Modulus [kPa]** | | | **Effects** | | |
|---|---|---|---|---|---|---|---|---|---|---|---|
| | (F/M) | (years) | | **Methods** | (Hz) | Glo/Dis | Gray | White | Age | Gender | Disease |
| Chan, et al. [197] | 35 (20/15) | 1-62 | BT | USE | 3-15e6 | | | | | | √ (BT) |
| Garcés Iñigo, et al. [145] | 57 (25/32) | neonates | T,CC | USE | 4-9e6 | | 1.17 | 1.60 | √ (0) | √ (0) | |
| McIlvain, et al. [140] | 125 (62/63) | 5-35 | WB,G,W,H, T | MRE | 50 | 3.17 | 3.20 | 3.22 | √ (-) | | |
| Burman Ingeberg, et al. [166] | 8 (3/5) | 21-33 | G,W | MRE | 50 | 0.21 | 0.20 | 0.22 | | | |
| Karki, et al. [175] | 137 | | | MRE | 60 | | | | | | √ (NPH) |
| Klemmer Chandía, et al. [146] | 10 (9/1) | 25-40 | C | USE | 27-56 | | 1.30 | | √ (-) | | |
| Triolo, et al. [167] | 18 (9/9) | 24-31 | WB,G,W,C, H,T,CC | MRE | 50 | 2.73 | 2.70 | 2.84 | | | |
| Yu, et al. [194] | 1 (1/0) | -- | BT | USE | 180-300 | 1.47/2.37 | | | | | √ (BT) |

Similar to findings from invasive testing, studies using MRE and USE have consistently reported frequency-dependent stiffening behavior in brain tissue [137, 138, 161, 162]. However, results on age-related effects exhibit substantial inter-study variability. For example, Sack and their colleagues observed a pronounced age-related decline in brain stiffness, reporting a reduction rate of approximately 0.75 % to 0.8 % per year [159, 161]. Their findings were supported by McIlvain, et al. [140] and Klemmer Chandía, et al. [146] in their brain aging analyses using MRE and USE, respectively. In contrast, several USE-based studies have reported positive maturation effects, suggesting increased brain elasticity with age [145, 191, 195], while others found no significant correlation between age and brain stiffness in MRE studies [138, 151, 162]. Gender-related differences in brain mechanical properties have also been explored. Some studies reported that female brains exhibit higher stiffness compared to male brains [159, 161, 162], while other investigations using USE found no significant discrepancy between sexes [145, 192]. Moreover, the distinction between healthy and pathological brain tissue has been examined across various neurological conditions. Altered mechanical properties have been observed in cases of brain tumors [158, 169, 194, 196, 197], AD [172], epilepsy [176], PD [173], dementia [174] and normal pressure hydrocephalus [175, 200].

An interesting observation from noninvasive testing methods is the finding that white matter appears consistently stiffer than gray matter in both MRE and USE measurements. In contrast, invasive testing methods show mixed trends in brain shear stiffness, as shown in Figure 4a. To provide a comparative overview, we compile the shear stiffness values obtained from both invasive and noninvasive approaches, alongside the instantaneous shear stiffness values for reference, as shown in Figure 9. Noninvasive measurements generally yield higher shear stiffness values compared to invasive methods (Figure 9a), and these values tend to fall within the range of instantaneous shear stiffness (Figure 9b). This outcome is largely expected and can be attributed to several factors. Invasive methods such as AMT, OST, or IND typically rely on post-conditioning data for parameter calibration. These values are naturally lower than preconditioning responses due to potential microstructural reorganization or redistribution of interstitial fluid, as discussed in Section 3.1. In this context, preconditioning data may more closely resemble *in vivo* mechanical behavior [125]. Additionally, tissue degradation in post-mortem samples further contributes to the lower shear stiffness reported in *ex vivo* studies. This contrast between invasive and noninvasive results ties into the broader discussion of *in vivo* versus *ex vivo* (or *in vitro*) mechanical characterization, which is an ongoing debate explored in numerous studies [21, 43, 45, 221].

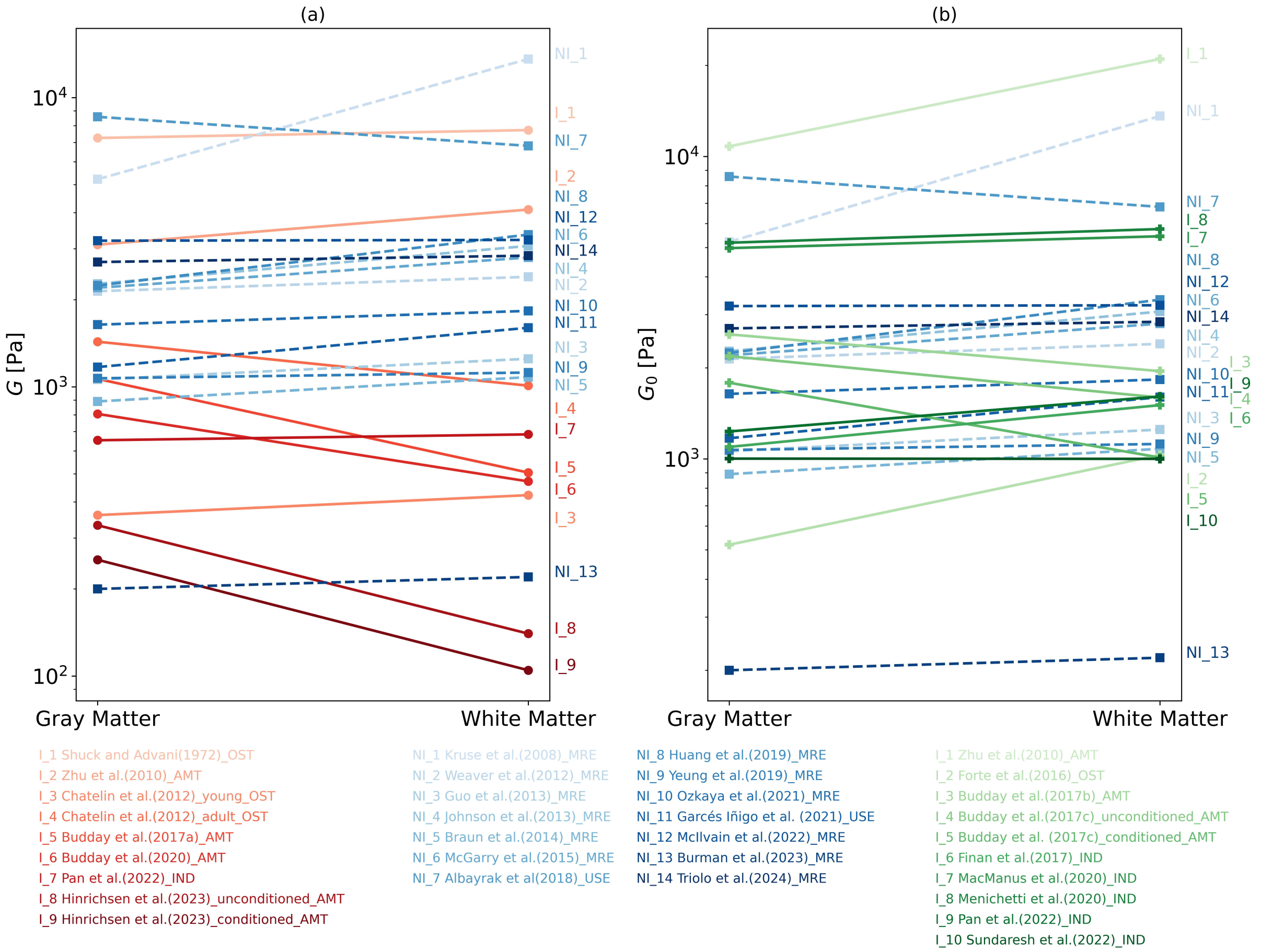

**Figure 9. Comparison of gray matter and white matter shear properties from both invasive and non-invasive studies**. Data compiled from Tables 2 and 3. (a) Shear modulus ($G$) from invasive and noninvasive measurements; (b). Instantaneous shear modulus ($G_0$) from invasive studies compared to shear modulus from noninvasive studies. For studies reporting multiple regions within gray or white matter, modulus values are averaged across regions. For studies with measurements at multiple frequencies, only the modulus at the lowest frequency is shown.

However, inter-study comparisons of human brain mechanics are extremely limited due to ethical considerations that make invasive *in vivo* testing on human subjects both difficult and controversial. A rare example is the pilot study by Schiavone, et al. [47], who introduced a light-based aspiration device for post-operative *in vivo* elasticity measurements. However, this method was restricted to shallow cortical indentation (1-3 mm) of cortex and suffered from measurements artifacts due to poor synchronization between applied pressure and imaging-based deformation tracking. Conversely, the reliability of intra-study comparison is also limited by methodological inconsistencies, including difference in sample characteristics (e.g., subject variation, anatomical location, sample dimension), testing conditions (e.g., apparatus, temperature, humidity, loading rate), and data interpretation approaches (e.g., model assumptions, calibration techniques). Noninvasive results are further affected by factors such as actuation frequency, operator expertise, imaging quality, and data post-processing. As such, it remains a major challenge to determine which testing methods best reflects the actual mechanical behavior of human brain tissue or quantify the discrepancy between *in vivo* and *ex vivo* results presuming the former as a golden standard. An alternative approach is to use the animal brains for comparative analyses. Table 4 provides a summary of selected animal studies to offer insights into the differences between *in vivo* and *ex vivo* testing conditions. Although animal studies provide valuable insights and controlled environments for experimentation, caution must be exercised when extrapolating their results to human brain mechanics, particularly in the context of clinical relevance and model calibration.

**Table 4. Summary of animal brain studies conducted *in vivo*, *in situ*, *in vitro*, and *ex vivo*.** PMI: post-mortem interval. IND: indentation, OST: oscillatory shear test, MRE: magnetic resonance elastography, USE: ultrasound elastography.

| Literature | Animals | Methods | | | | | Findings |
|---|---|---|---|---|---|---|---|
| | | *in vivo* | *in situ* | *in vitro* | *ex vivo* | PMI | |
| Fallenstein, et al. [206] | Monkey | OST | | OST | | 2-5h | • No significant difference between loss tangent *in vivo* and *in vitro*. |
| Gefen, et al. [222] | Porcine | IND | IND | IND | | 6h | • Long-term time constant of relaxation significantly decreased form *in vivo* to *in situ* modes;<br>• Preconditioning decreased the shear moduli, with a more pronounced effect *in situ* and *in vitro*. |
| Vappou, et al. [223] | Rat | MRE | | | MRE | 0.5-24h | • Significant increase in shear storage modulus of about 100% was found to occur just after death;<br>• Insignificant difference between shear loss moduli *in vivo* and *ex vivo* (0.5h), and a decrease of about 50% was found to occur after 24h. |
| Prevost, et al. [85] | Porcine | IND | IND | IND | | 6-7h | • The indentation response was significantly stiffer *in situ* than *in vivo* by a factor of 1.5–2;<br>• The indentation response *in vitro* was more compliant than *in situ*, with peak forces 20% lower *in vitro*. |
| Urbanczyk, et al. [224] | Porcine | USE | USE | USE | | 4-5h | • Shear modulus *in situ* and *in vitro* were 37% and 22% higher than *in vivo* moduli;<br>• Brain stiffness decreases with increased temperature (23%) and external confinement (22–37%). |
| Guertler, et al. [225] | Porcine | MRE | | | MRE | 2h | • Brain tissue *in vivo* appears stiffer than *ex vivo* at frequencies of 100 Hz and 125 Hz;<br>• Brain mechanical difference between *in vivo* and *ex vivo* becomes smaller at lower frequencies. |
| Liu, et al. [143] | Rabbit | USE | | | USE | 1h | • Shear modulus from *in vivo* measurements is about 47% higher than *ex vivo* measurements;<br>• The change in *ex vivo* elastic properties within 60-min post-mortem is negligible. |

## 4. Summary, Challenges, and Perspectives

As the most functionally complex and vital organ in the human body, the brain exhibits highly intricate and unique mechanical behaviors. Understanding these properties is essential not only for deepening our knowledge of fundamental brain physiology but also for informing clinical applications such as surgical planning, trauma modeling, and disease diagnosis. However, accurately characterizing brain mechanics remains exceptionally challenging due to their biphasic composition, extreme softness and fragility, and the structural heterogeneity arising from its diverse cellular populations and anatomically distinct subregions. In this review, we systematically introduced the most commonly used mechanical testing techniques applied to brain tissue, including both invasive approaches (AFM, IND, AMT, and OST) and noninvasive modalities (MRE and USE). Each of these techniques offers specific advantages and faces particular limitations, and their applicability is often dictated by distinct spatial and temporal requirements, as well as the experimental context. For example, AFM and IND are well-suited for probing localized mechanical properties at multiple resolutions, ranging from the cellular or subcellular to the tissue level, but their testing accuracies are sensitive to the geometry of the indenter tip and the mechanical models for characterization. AMT and OST provide versatile platforms for assessing regional mechanical properties and capturing nonlinear, anisotropic behaviors, yet often require complex sample preparations and are susceptible to boundary effects induced by tissue fixation. On the other hand, MRE and USE allow for noninvasive, *in vivo* mapping of brain mechanics, enabling large-cohort studies and regional comparisons. Nonetheless, their operation is typically limited to small deformations, constrained by model assumptions and signal attenuation, especially imaging deep or structurally complex brain regions. By summarizing these techniques, we aim to provide a practical reference for researchers and clinicians in selecting the appropriate tools for investigating brain biomechanics across different application domains.

Recognizing the limitations of animal models due to species-specific differences, we further reviewed existing publications focused on examining the mechanical properties of human brain tissue. These studies were presented separately according to invasive and noninvasive approaches. We systematically summarized and compared the reported material properties, testing conditions, and parameter calibration strategies across all the studies reviewed. Invasive methods revealed a broad spectrum of mechanical properties, with various shear-related moduli values spanning several orders of magnitude. This variability is largely attributed to differences in tissue preparation, environmental testing conditions (e.g., temperature and humidity), and preconditioning effects, as well as post-mortem degradation in *ex vivo* samples. Noninvasive methods, in contrast, facilitated population-level assessments in terms of age, gender, and pathological conditions. These studies also revealed more consistent trends, such as higher stiffness observed in white matter relative to gray matter, although their spatial resolution and interpretability remain constrained by model-based assumptions. Furthermore, we found that the choice of constitutive modeling frameworks—ranging from linear elastic and hyperelastic to viscoelastic and poroelastic—also influences how mechanical properties are interpreted and reported. Based on our review and analysis, we suggest the following perspectives as potential considerations for advancing future studies on human brain tissue mechanical experiments:

***Standardized and clearly reported testing conditions are critical.***

In mechanical testing of brain tissue, it is imperative to report sampling information and experimental settings with clarity and precision. This includes, but is not limited to, (1) subject-related details: age, gender, and any relevant pathological conditions; (2) critical sampling parameters for invasive testing: post-mortem interval, anatomical locations and orientation of the extracted tissue, preservation method, specimen's geometry and dimensions, humidity control, and testing temperature; (3) loading conditions: testing apparatus, preconditioning, loading rate or frequency, the deformation ranges, and method of data acquisition; (4) model assumption: compressibility, mechanical simplifications, and the choice of constitutive models. All of these factors can influence the measured outcomes of brain tissue. Comprehensive and transparent reporting of this information not only enhances the reproducibility of experimental findings but also facilitates meaningful intra- and inter-study comparisons, offering valuable guidance for future research in the field.

***Noninvasive methods are promising but cannot fully replace invasive techniques.***

Noninvasive techniques such as MRE and USE provide powerful capabilities for measuring brain mechanical properties *in vivo*. These methods enable large-scale studies and longitudinal assessments without the need for tissue extraction. However, their accuracy is highly sensitive to the imaging resolution, operator experience, and efficiency of shear wave excitation mechanisms. Both MRE and USE operate in the small-strain regime, and therefore restricting their ability to capture the nonlinear responses that emerge under large deformations, which are particularly relevant in modeling brain mechanics. Additionally, the calibration of mechanical parameters often relies on simplifying assumptions and predefined mechanical models, which may not fully capture the brain's complex mechanical responses. Invasive methods, though facing ethical constraints in human studies, remain indispensable for obtaining high-fidelity data, particularly in cases requiring fine-scale or multi-modal testing. Moreover, the invasive methods enable direct measurement of continuous force-displacement or stress-strain relationships under controlled loading conditions, offering critical insights into brain mechanical behavior and serving as a foundation for robust material model development. Albeit with current limitations, elastography-based noninvasive methods still hold great promise in transforming the measurement paradigm. With ongoing advancements in imaging technology, artificial intelligence (AI), including machine learning (ML) and deep learning (DL), the noninvasive methods are expected to become increasingly accurate and reliable, thereby narrowing the gap that currently exists between practicality and precision in brain biomechanics.

***Multiscale and multimodal testing should be encouraged to capture the full mechanical landscape.***

Brain tissue exhibits distinct mechanical behaviors across spatial and temporal scales. Techniques such as AFM enable the probing of nano- or microscale stiffness heterogeneity, while AMT provide insight into bulk tissue responses. Integrating data from multiple modalities (e.g., AFM, IND, AMT, MRE) across different scales allows for a more comprehensive understanding of how microstructure features influence overall tissue mechanics, thereby improving the fidelity of biomechanical models [114, 226]. To date, no single constitutive model with a unified parameter set can fully capture the complex mechanical responses of brain tissue under all loading scenarios. Each deformation mode—whether tension, compression, or shear—reveals unique aspects of the tissue's behavior. Therefore, incorporating multimodal experimental data into mechanical characterization is crucial for enhancing the robustness, generalizability, and predictive accuracy of constitutive models. Such integrative approaches are essential for building a more complete and realistic representation of brain tissue mechanics.

***Inverse modeling with FEM offers a more accurate and physically reliable characterization of material parameters.***

Given the complex deformation patterns and boundary conditions involved in brain mechanics, inverse parameter identification methods, especially those using FEM, are essential for accurately extracting material properties [227]. These approaches allow researchers to go beyond simple curve fitting and simulate the actual testing environment, effectively reducing artifacts introduced by the boundary effects, sample geometry inconsistencies, or deformation caused by gravitational force [25, 93]. Conventional FEM-based inverse identification often presumes a specific material model, such as hyperelastic, viscoelastic, poroelastic, or combined forms, followed by iterative optimization of the model parameters. This is done by minimizing the discrepancy between simulated mechanical responses (deformation or stress) and experimental observations [228]. While this strategy generally yields more reliable parameter calibration than direct fitting of experimental data, it is often computationally expensive and time-consuming regarding FEM modeling. Moreover, the reliance on predefined material model forms inherently constrains the discovery of novel constitutive behaviors or unanticipated mechanical features present in experimental data. Emerging approaches based on ML, including data-driven inverse modeling and automated parameter discovery framework, have shown promise in overcoming these limitations [229-231].

***While precision matters, relative trends often suffice for clinical applications.***

Brain tissue is composed of a diverse array of living cells, each contributing to its structurally and functionally heterogeneous nature. Given its dynamic and evolving properties—shaped by factors like age, microstructural remodeling, and disease progression—achieving a universally "accurate" mechanical parameter is often unrealistic. For many clinical applications, such as disease diagnosis or monitoring, it is more meaningful to assess relative changes in tissue stiffness, spatial gradients, and temporal trends, or propose a safe physiological range, rather than relying solely on absolute material constants. As a result, *in vivo* techniques like USE are more frequently applied in clinical practice than in fundamental research. However, accurate and standardized testing protocols are also needed considering the wide variation in reported mechanical parameters (often spanning several orders of magnitude), especially in invasive testing.

Looking forward, we hope this review will serve as a valuable resource for advancing the field of brain biomechanics. We encourage future research to bridge the gap between *in vivo* and *ex vivo* findings, standardize testing protocols, and develop more physiologically relevant models that capture the complex, nonlinear, and time-dependent behavior of human brain tissue.

## 5. Acknowledgement

JH, KJ, TL, and XW acknowledges the support from the National Science Foundation (IIS-2011369) and National Institutes of Health (1R01NS135574-01). KS acknowledges the support from Women and Philanthropy Foundation (N/A), Arizona Biomedical Research Centre (award # RFGA2022-010-07), 2023 Mayo Clinic and Arizona State University Alliance for Health Care Collaborative Research Seed Grant (N/A), and the NSF Faculty Early Career Development Program (CAREER) award (2145895). GL acknowledges the support from the National Institutes of Health (EB037388), MJR acknowledges the support from the National Science Foundation (CMMI: 2123061).

## 6. Data Availability Statement

No data was used for the research described in this manuscript.

## 7. Conflict of Interest

The authors declare that the research was conducted in the absence of any commercial or financial relationships that could be construed as a potential conflict of interest.

## 8. Authorship Contribution Statement

JH, KJ: Methodology, Conceptualization, Investigation, Writing – Original Draft; AR, AK, WZ, LZ, TW, RP, DZ, GL, KS, TL, MR, EK: Writing – Review and Editing; XW: Conceptualization, Supervision, Funding acquisition, Writing – Original Draft, Writing – Review and Editing.